\documentclass[global]{svjour}
\usepackage{amsfonts}
\usepackage{amsmath}

\input{tcilatex}

\begin{document}

\journalname{Submitted to ...}
\title{\textbf{GEOMETRY OF INTERACTIONS IN COMPLEX BODIES}}
\author{Chiara de Fabritiis\inst{1} \and Paolo Maria Mariano\inst{2}}
\institute{Dipartimento di Scienze Matematiche,\\
Universit\`{a} Politecnica delle Marche,\\
via Brecce Bianche, Ancona (Italy)\\
\email{fabritii@dipmat.univpm.it}%
\and Dipartimento di Ingegneria Strutturale e Geotecnica,\\
Universit\`{a} di Roma "La Sapienza",\\
via Eudossiana 18, I-00184 Roma (Italy)\\
\email{paolo.mariano@uniroma1.it}%
,\\
now at Universit\`{a} del Molise, Campobasso (Italy).}
\maketitle

\begin{abstract}
We analyze geometrical structures necessary to represent bulk and surface
interactions of standard and substructural nature in complex bodies. Our
attention is mainly focused on the influence of diffuse interfaces on sharp
discontinuity surfaces. In analyzing this phenomenon, we prove the
covariance of surface balances of standard and substructural interactions.
\end{abstract}


\section{Introduction}

Bodies are called \emph{complex} when their material substructure (i.e. the
texture from nano-level to meso-level) has a prominent influence on their
gross behavior. Such an influence appears also through a not negligible
occurrence of interactions due to substructural changes. Examples are liquid
crystals, elastomers, ferroelectric and microcracked bodies, spin glasses.
Above all, soft condensed matter displays complex behavior. Applications in
nanotechnology, smart structures and various fields of technology open basic
theoretical and experimental problems that challenge, in a certain sense,
even some aspects of the foundational concepts of standard continuum
mechanics. Basically, the standard paradigm of Cauchy's format of continuum
mechanics, prescribing that the material element is a sort of indistinct
sphere that we collapse in a point in space seems to be not sufficient to
account for the articulated substructural nature of a complex body.

In fact, for complex bodies the material element is rather a `system' and
one needs the introduction of appropriate morphological descriptors $\mathbf{%
\nu }$ of such a system (order parameters) at least at a coarse grained
level. They describe the essential geometrical features of substructural
shapes.

Physical circumstances of disparate nature suggest a rich crop of possible
choices of $\mathbf{\nu }$, each one characterizing special models.
Moreover, the selection of morphological descriptors is strongly related
with the representation of substructural interactions arising within each
material element and between neighboring material elements as a consequence
of substructural changes. Interactions are represented in fact by objects
conjugated in the sense of power with the rates of the quantities describing
the geometry of the body and its changes. In this sense, since placement and
order parameter fields are involved, the description of complex bodies
adopted here is called \emph{multifield}. It has basic differences with
standard internal variable models. In a multifield approach, morphological
descriptors enter directly the geometrical representation of the body and
its kinematics; true interactions are associated with their rates and
balanced. Information about the material substructure are introduced already
at the level of geometrical description of the body. On the contrary, in
standard internal variable models, the geometrical description of the body
is of Cauchy's type: the material element is morphologically equivalent to
an indistinct sphere described just by its place in space. There, internal
variables come into play in constructing thermodynamics to carry information
about the material substructure just at constitutive level (a detailed
treatment of these classes of models can be found in [34]).

By following the general unified framework of multifield theories proposed
by Capriz in 1989 [4] (see also [3]) and then developed further on in its
abstract structure (see [32], [23], [8], [9]), we do not specify the nature
of $\mathbf{\nu }$. We require only that $\mathbf{\nu }$ be an element of a
finite-dimensional differentiable paracompact manifold without boundary $%
\mathcal{M}$ to cover a class as large as possible of special theories. Our
attention is then focused on the general setting which contains as special
cases prominent theories interpreting problems typical of condensed matter.
In a certain sense our work deals with a model of models. However, if we
tackle this general point of view, we face the basic difficulty that $%
\mathcal{M}$\ does not coincide with a linear space in general. Moreover, we
cannot consider a priori $\mathcal{M}$\ embedded in a linear space because
the embedding is a non-trivial part of some special modeling.

We focus our attention on conservative processes. For them, the relevant
appropriate Hamiltonian formalism has been developed in [9] as an evolution
of the Hamiltonian formalism in classical non-linear elasticity (the one
discussed in [27]). We start from the results in [9] and analyze some of the
rather subtle geometrical questions induced by the abstract nature of $%
\mathcal{M}$.

Our essential point of view is the one of [25]: "Geometry and mechanics of
maps between manifolds are a general framework for condensed matter physics
and are also a tool to construct new models of unusual and perhaps
unaspected phenomena."

The main attention is focused on the representation of the way in which
diffuse interfaces influence the behavior of additional sharp interfaces. In
fact, the presence of the gradient of $\mathbf{\nu }$ in the list of entries
of the energy may account for diffuse interfaces as well as it may describe
weakly non-local (i.e. gradient) effects in homophase materials. Additional
discontinuity surfaces may also occur as a consequence of the presence of
defects, shock or acceleration waves, solidification phenomena of complex
fluids etc. As an example one may consider a polarized ferroelectric
material in which external loads induce a shock wave: the shock front
encounters walls of polarized domains and interact with them. So, in
general, one may say that, in principle, diffuse interfaces and potential
homophase gradient effects may interact with sharp discontinuity surfaces
influencing their possible evolution. In particular, when discontinuity
surfaces are structured, in the sense that they are endowed with surface
energy, surface interactions may occur. They are of standard and
substructural nature. Standard surface stresses are in fact due to the
deformation of the discontinuity surface while the other are a consequence
of changes of substructure at the surface itself.

Surface balances involving standard surface stresses have been obtained in
[17] while substructural measures of interactions have been introduced in
[22], [23] and the relevant balance equations derived there (see also [8]).
However, there is no proof of their covariance. Such a proof is provided
here (see Theorem 2) and is the main result of this paper. In obtaining it
we enlarge the standard notion of observer: for us, in fact, such a notion
involves not only the representation of the ambient space and the time
scale, but also the representation of the manifold of substructural
morphologies (see also [25]).

\section{Morphology of complex bodies}

We consider a body occupying a regular region $\mathcal{B}_{0}$ of the
three-dimensional Euclidean space $\mathcal{E}^{3}$ (with affine translation
space $Vec$)\footnote{%
With the adjective `regular' we refer to an open (bounded) subset of $%
\mathcal{E}^{3}$ coinciding with the interior of its closure, with a
surface-like boundary where the outward unit normal $\mathbf{n}$ is
well-defined everywhere to within a finite number of corners and edges. The
treatment of infinite bodies requires only some minor technical adjustments
in the results of the present paper. In any case, some remarks about them
are presented throughout the paper.}. The current morphology of the body is
described by two sufficiently smooth mappings:%
\begin{equation}
\mathcal{B}_{0}\mathcal{\ni }\mathbf{X}\overset{\mathbf{\tilde{x}}}{\mathbf{%
\longmapsto }}\mathbf{x=\tilde{x}}\left( \mathbf{X}\right) \in \mathcal{E}%
^{3}\text{ \ \ \ , \ \ \ \ }\mathcal{B}_{0}\mathcal{\ni }\mathbf{X}\overset{%
\mathbf{\tilde{\nu}}}{\mathbf{\longmapsto }}\mathbf{\nu =\tilde{\nu}}\left( 
\mathbf{X}\right) \in \mathcal{M}.  \label{1}
\end{equation}

\begin{enumerate}
\item $\mathbf{\tilde{x}}$ shows the current \emph{placement} of a material
element at $\mathbf{X}$ in $\mathcal{B}_{0}$, is injective and orientation
preserving, and $\mathcal{B}=\mathbf{x}\left( \mathcal{B}_{0}\right) $ is
also regular as $\mathcal{B}_{0}$.

\item $\mathbf{\tilde{\nu}}$ is the \emph{order parameter }map and assigns
to each material element a coarse grained morphological descriptor $\mathbf{%
\tilde{\nu}}\left( \mathbf{X}\right) $\ of its substructure (order
parameter), chosen as an element of a differentiable paracompact manifold $%
\mathcal{M}$ (generally without boundary).
\end{enumerate}

Two natural tangent maps arise, namely $T\mathbf{x}:T\mathcal{B}%
_{0}\rightarrow T\mathcal{B}$ and $T\mathbf{\nu }:T\mathcal{B}%
_{0}\rightarrow T\mathcal{M}$. The pairs $\left( \mathbf{x},\nabla \mathbf{x}%
\right) $, $\left( \mathbf{\nu },\nabla \mathbf{\nu }\right) $ are the
peculiar elements of $T\mathbf{x}$ and $T\mathbf{\nu }$ respectively. Since $%
T\mathcal{B}_{0}$\ is a trivial bundle and a connection is natural over it,
we can separate in invariant way $\mathbf{x}$ from $\nabla \mathbf{x}$ which
is commonly indicated with $\mathbf{F}$. The condition that $\mathbf{\tilde{x%
}}$ be orientation preserving implies that, at each $\mathbf{X}\in \mathcal{B%
}_{0}$, $\mathbf{F}\in Hom\left( T_{\mathbf{X}}\mathcal{B}_{0},T_{\mathbf{%
\tilde{x}}\left( \mathbf{X}\right) }\mathcal{B}\right) $ \emph{has positive
determinant}. Since $\mathcal{M}$ is a priori not trivial, the pair $\left( 
\mathbf{\nu },\nabla \mathbf{\nu }\right) $ cannot be separated in invariant
way, unless there is a parallelism over $\mathcal{M}$. In principle, one may
define in abstract way a parallelism over $\mathcal{M}$ but, since the pair $%
\left( \mathbf{\nu },\nabla \mathbf{\nu }\right) $ enters constitutive
issues, one should have a physically significant parallelism. In other
words, when we act separating in invariant way $\nabla \mathbf{\nu }$ from $%
\mathbf{\nu }$ we should presume to have at least one physically significant
parallelism over $\mathcal{M}$ even when circumstances may allow us the use
of invariance requirements with respect to the choice of the connection. In
fact, the presence of the pair $\left( \mathbf{\nu },\nabla \mathbf{\nu }%
\right) $\ or of arbitrarily one element of it (namely $\mathbf{\nu }$\ or $%
\nabla \mathbf{\nu }$) in the list of entries of the energy changes the
representation of interactions.

As remarked in the introduction, in dealing with the geometrical
(morphological) description of complex bodies, we relax one of the axioms of
the mechanics of simple materials. In our picture, in fact, the material
element is not morphologically equivalent to a `monad', a simple material
particle in the sense of Cauchy (see [29], [30]), identified only by its
place in space. Instead, we consider the body as a collection of \emph{%
subsystems} of the \emph{same} nature (the material substructure) and the
order parameter at a given point represents the characteristic features of
the morphology of the subsystem there.

Sometimes the material substructure is a perfectly identifiable Lagrangian
system as in the case of nematic liquid crystals [14], [20], [11] (in which
the nematic molecules can be separated from the melt), sometimes it does not
as in granular gases [6] and microcracked bodies [26]. In granular gases,
e.g., a material element collects a family of sparse granules with peculiar
velocities, so that the order parameter may be an element of a suitable
Grassmanian of the tangent bundle of some finite-dimensional manifold, while
for microcracked bodies each microcrack can be considered either as a sharp
planar defect not interpenetrated by interatomic bonds or as an elliptic
void, so it does not exist \emph{per se} and has a volatile substance
determined just by the surrounding matter.

We leave undetermined the specific nature of $\mathbf{\nu }$\ to cover a
class as large as possible of special cases, following in this way the
unifying point of view of Capriz [4]. Our primary strategy is to work with
minimal requirements for $\mathcal{M}$ and to add geometrical structure to
it only when necessary.

In this way we avoid the path leading to a monotonous zoo of special models
with slightly different formal aspects but with the same intrinsic nature.

With respect to Hamiltonian elasticity of simple bodies (see [27], Chap. 4),
here the main source of difficulties is the circumstance that $\mathcal{M}$
is a non-trivial manifold; in particular \emph{it does not coincide} with a
linear space in most cases of prominent interest.

The space $\mathcal{C}$ of pairs of maps $\left( \mathbf{\tilde{x},\tilde{\nu%
}}\right) $, a product space of the type $\mathcal{C}_{\mathbf{x}}\times 
\mathcal{C}_{\mathbf{\nu }}$, with $\mathbf{\tilde{x}}$ pertaining to $%
\mathcal{C}_{\mathbf{x}}$\ and $\mathbf{\tilde{\nu}}$ to $\mathcal{C}_{%
\mathbf{\nu }}$, has a non-trivial structure which depends on the
geometrical properties of $\mathcal{M}$. Basically, we imagine that $%
\mathcal{C}_{\mathbf{x}}\subseteq W^{1,p}\left( \mathcal{B}_{0},Vec\right) $
for some $p\geq 1$ and $\mathcal{C}_{\mathbf{\nu }}=PC^{1}\left( \mathcal{B}%
_{0},\mathcal{M}\right) $, i.e. we require that the order parameter map be
at least continuous and piecewise continuously differentiable over $\mathcal{%
B}_{0}$, while the placement map be an element of the Sobolev space $%
W^{1,p}\left( \mathcal{B}_{0},Vec\right) $, even if we may basically require
that $\mathbf{\tilde{x}}$ be continuous and piecewise continuously
differentiable too.

Specific examples showing the possible intricate nature of $\mathcal{C}$, in
particular of $\mathcal{C}_{\mathbf{\nu }}$, can be discussed at length.

We analyze below the case in which $\mathcal{M}$ has a Riemannian structure.
However, the developments in subsequent sections do not require strictly
such a structure that will be called upon only when necessary.

In any case, even when $\mathcal{M}$ is Riemannian, we do not assign a
priori any prevalent r\^{o}le to Levi-Civita connection.

\subsection{The case in which $\mathcal{M}$ has a Riemannian structure:
aspects of the nature of $\mathcal{C}$}

We assume just in this section that $\mathcal{M}$ is a Riemannian,
connected, finite dimensional manifold, endowed with a metric $\mathbf{g}_{%
\mathcal{M}}$ and a consequent Levi-Civita connection.

Notice that \emph{we do not require here the embedding of} $\mathcal{M}$ 
\emph{in some linear space}. When, in fact, we embed $\mathcal{M}$ in a
linear space by using, let say, Nash's isometric embedding to preserve at
least he quadratic part of the kinetic energy (if it exists) pertaining to
the substructure, the embedding is not unique. So the choice of it becomes
matter of modeling. Moreover, the embedding is also not strictly necessary
to build up the structures needed for the basic aspects of mechanics. For
this reason, with the aim to eliminate overstructures as much as possible,
we consider $\mathcal{M}$ in its abstract setting, not embedded a priori in
a linear space.

We denote with $\left\langle \cdot ,\cdot \right\rangle _{T\mathcal{M}}$ the
scalar product over $T_{\mathbf{\nu }}\mathcal{M}$ associated with $\mathbf{g%
}_{\mathcal{M}}$. For any $C^{1}$-curve $\left[ 0,s^{\ast }\right] \ni
s\longmapsto \mathbf{\nu }\left( s\right) \in \mathcal{M}$, we have a vector
field $\mathbf{\upsilon }\left( s\right) =\frac{d\mathbf{\nu }}{ds}\left(
s\right) $, and for any other vector field $A\left( s\right) $ over $%
\mathcal{M}$ one defines its derivative along $s\longmapsto \mathbf{\nu }%
\left( s\right) $ writing $\frac{DA}{ds}=\nabla _{\mathbf{\upsilon }\left(
s\right) }A$, and call $A$ \emph{autoparallel along} $s\longmapsto \mathbf{%
\nu }\left( s\right) $ when $\frac{DA}{ds}=0$ for any $s\in \left[ 0,s^{\ast
}\right] $. As usual, $s\longmapsto \mathbf{\nu }\left( s\right) $ is a 
\emph{geodesic} when $s\longmapsto \mathbf{\upsilon }\left( s\right) $ is
autoparallel along it.

The Riemannian structure assures the existence of a natural distance $d_{%
\mathcal{M}}:\mathcal{M}\times \mathcal{M}\rightarrow \mathbb{R}^{+}$ over $%
\mathcal{M}$. In fact, by indicating with $\lambda $ any arbitrary piecewise 
$C^{1}$ curve $\lambda :\left[ 0,s^{\ast }\right] \rightarrow \mathcal{M}$
such that $\lambda \left( 0\right) =\mathbf{\nu }_{1}$ and $\lambda \left(
s^{\ast }\right) =\mathbf{\nu }_{2}$, as usual we put 
\begin{equation}
d_{\mathcal{M}}\left( \mathbf{\nu }_{1},\mathbf{\nu }_{2}\right) =\inf
\left\{ l\left( \lambda \right) \right\} ,  \label{2}
\end{equation}%
varying $\lambda $ between $\mathbf{\nu }_{1}$ and $\mathbf{\nu }_{2}$, with 
$l\left( \lambda \right) $ the length%
\begin{equation}
l\left( \lambda \right) =\int_{0}^{s^{\ast }}\left\langle \frac{d\lambda }{ds%
}\left( s\right) ,\frac{d\lambda }{ds}\left( s\right) \right\rangle _{T%
\mathcal{M}}^{\frac{1}{2}}ds  \label{3}
\end{equation}%
(see, e.g., [1] Chapter 2, [33] Chapter 4 ). Locally, $d_{\mathcal{M}}$\ is
calculated over geodesics.

However, $d_{\mathcal{M}}$ may be unbounded. In this case, we select a new
metric $\tilde{d}_{\mathcal{M}}:\mathcal{M}\times \mathcal{M}\rightarrow 
\mathbb{R}^{+}$ over $\mathcal{M}$ defined by%
\begin{equation}
\tilde{d}_{\mathcal{M}}=\frac{d_{\mathcal{M}}}{1+d_{\mathcal{M}}}.  \label{4}
\end{equation}

Notice that the metric $\tilde{d}_{\mathcal{M}}$

\begin{description}
\item[(a)] is equivalent to $d_{\mathcal{M}}$\ (it leads to the same
topology),

\item[(b)] is bounded,

\item[(c)] is of class $C^{\infty }\left( \mathcal{M}\right) $,

\item[(d)] if $d_{\mathcal{M}}$ is complete on $\mathcal{M}$, then $\tilde{d}%
_{\mathcal{M}}$ is complete too.
\end{description}

A well know theorem due to Hopf and Rinow ([19], p. 56) asserts that the
following statements are equivalent: (i) $d_{\mathcal{M}}$ is complete, (ii)
closed bounded sets with respect to $d_{\mathcal{M}}$ are compact, (iii)
geodesics for $d_{\mathcal{M}}$ can be continued as to be defined on the
whole real line. Moreover, if $\mathcal{M}$ is complete, the distance $d_{%
\mathcal{M}}$\ can be computed over geodesics not only locally, but also
globally.

\textbf{Remark 1}. When $\mathcal{M}$ is compact, $d_{\mathcal{M}}:\mathcal{M%
}\times \mathcal{M}\rightarrow \mathbb{R}^{+}$ is bounded, then it is not
necessary to substitute it with $\tilde{d}_{\mathcal{M}}$. Of course, the
definition of $\tilde{d}_{\mathcal{M}}$ is just one of the possible ways to
bound the metric (i.e., the `diameter' of the manifold is finite).

By making use of $\tilde{d}_{\mathcal{M}}$\ we may define three natural
distances over $\mathcal{C}_{\mathbf{\nu }}$.

\begin{description}
\item[1.] The first metric distance, indicated with $d^{\left( i\right) }$,\
is defined by%
\begin{equation}
d^{\left( i\right) }\left( \mathbf{\tilde{\nu}}_{1},\mathbf{\tilde{\nu}}%
_{2}\right) =\int_{\mathcal{B}_{0}}\tilde{d}_{\mathcal{M}}\left( \mathbf{\nu 
}_{1},\mathbf{\nu }_{2}\right) d\left( vol\right) ,  \label{5}
\end{equation}%
where (\emph{i}) stands for `integral'.

\item[2.] Let $\left\{ K_{n}\right\} $ be an exhaustion of $\mathcal{B}_{0}$%
, i.e. a compact cover of $\mathcal{B}_{0}$ such that $K_{n}\subset \overset{%
\circ }{K}_{n+1}$ for any $n$. The second distance, indicated with $%
d^{\left( c\right) }$,\ is defined by%
\begin{equation}
d^{\left( c\right) }\left( \mathbf{\tilde{\nu}}_{1},\mathbf{\tilde{\nu}}%
_{2}\right) =\dsum\limits_{n\in \mathbb{N}}\frac{1}{2^{-n}}\max_{\mathbf{%
X\in }K_{n}}\tilde{d}_{\mathcal{M}}\left( \mathbf{\nu }_{1},\mathbf{\nu }%
_{2}\right) ,  \label{6}
\end{equation}%
where (\emph{c}) stands for `compact'.

\item[3.] More simply, a third natural distance, indicated with $d^{\left(
s\right) }$,\ can be defined by%
\begin{equation}
d^{\left( s\right) }\left( \mathbf{\tilde{\nu}}_{1},\mathbf{\tilde{\nu}}%
_{2}\right) =\sup_{\mathbf{X\in }\mathcal{B}_{0}}\tilde{d}_{\mathcal{M}%
}\left( \mathbf{\nu }_{1},\mathbf{\nu }_{2}\right) ,  \label{7}
\end{equation}%
where (\emph{s}) stands for `supremum'.
\end{description}

\textbf{Remark 2}. There is a peculiar physical difference between the
metric $d^{\left( c\right) }$\ and the other two. In fact, the exhaustion of 
$\mathcal{B}_{0}$ by compact sets, used to define $d^{\left( c\right) }$,
implies that the values of the order parameter map over boundary points of $%
\mathcal{B}_{0}$ may contribute slightly to the distance in $\mathcal{C}_{%
\mathbf{\nu }}$, on the contrary of the other two metrics. In a certain
sense, $d^{\left( c\right) }$ seems to be preferable when there are
uncertainties in the physical meaning of boundary data about $\mathbf{\nu }$%
, a problem (the one of boundary data) that appears very subtle for some
material substructures like microcracks.

\textbf{Remark 3}. Notice that $d^{\left( c\right) }$\ and $d^{\left(
s\right) }$\ are complete over $C^{0}\left( \mathcal{B}_{0},\mathcal{M}%
\right) $ if $\tilde{d}_{\mathcal{M}}$ is complete. In particular,
compactness of $\mathcal{M}$ is sufficient for the validity of this
statement.

\textbf{Remark 4}. We realize that the space of continuous maps between $%
\mathcal{B}_{0}$ and $\mathcal{M}$, namely $C\left( \mathcal{B}_{0},\mathcal{%
M}\right) $, may be in general \emph{not complete} with respect to the
metric $d^{\left( i\right) }$. For example, let $\mathcal{B}_{0}$ coincide
with the cube $\left( -1,2\right) ^{3}$ in a frame $0X_{1}X_{2}X_{3}$. We
may analyze two different cases.

\begin{description}
\item[(i)] \emph{Case 1: }$\mathcal{M}$\emph{\ is not compact}. Let us
assume $\mathcal{M}\equiv \mathbb{R}$, and for any $x,y\in \mathbb{R}$, put $%
d_{\mathcal{M}}\left( x,y\right) =\left\vert x-y\right\vert $. Let us
construct also a sequence of $\mathbf{\nu }$ such that for $\left(
X_{2},X_{3}\right) $ ranging in $\left( -1,2\right) ^{2}$%
\begin{equation}
\mathbf{\nu }_{n}\equiv f_{n}\left( \mathbf{X}\right) =\left\{ 
\begin{array}{c}
0\text{ \ \ , \ \ }if\text{ }-1<X_{1}\leq 0 \\ 
X_{1}^{n}\text{ \ \ \ \ , \ \ \ }if\text{ }0\leq X_{1}\leq 1 \\ 
1\text{ \ \ \ \ , \ \ \ \ }if\text{ }1\leq X_{1}<2%
\end{array}%
\right. .  \label{8}
\end{equation}%
From (\ref{5}), for $n\leq m\in \mathbb{N}$, we get%
\begin{eqnarray}
d^{\left( i\right) }\left( f_{n},f_{m}\right) &=&\int_{\mathcal{B}_{0}}%
\tilde{d}_{\mathcal{M}}\left( f_{n}\left( \mathbf{X}\right) ,f_{m}\left( 
\mathbf{X}\right) \right) d\left( vol\right) \leq  \notag \\
&\leq &\int_{\mathcal{B}_{0}}d_{\mathcal{M}}\left( f_{n}\left( \mathbf{X}%
\right) ,f_{m}\left( \mathbf{X}\right) \right) d\left( vol\right) =  \notag
\\
&=&9\left( \frac{1}{n+1}-\frac{1}{m+1}\right) ,  \label{9}
\end{eqnarray}%
so that the sequence $f_{n}$ is Cauchy's. However, if we let $n$ to
infinity, the resulting limit is%
\begin{equation}
f\left( \mathbf{X}\right) =\left\{ 
\begin{array}{c}
0\text{ \ \ , \ \ }if\text{ }-1<X_{1}<1 \\ 
1\text{ \ \ \ \ , \ \ \ \ }if\text{ }1\leq X_{1}<2%
\end{array}%
\right. ,  \label{10}
\end{equation}%
which is not continuous. The physical situation covered by this example is
the one of a two-phase material in which the order parameter is the void
volume fraction of one of the two species. The limiting distribution (\ref%
{10}) describes the complete separation of the two species with the
occurrence of a discontinuity surface in the plane $X_{1}=1$.

\item[(ii)] \emph{Case 2: }$\mathcal{M}$\emph{\ is compact}. Let us assume $%
\mathcal{M}\equiv S^{1}\subset \mathbb{C}$ and define with a slight abuse of
notation%
\begin{equation}
\mathbf{\nu }_{n}\equiv f_{n}\left( \mathbf{X}\right) =\left\{ 
\begin{array}{c}
0\text{ \ \ \ \ \ \ \ \ , \ \ \ \ \ \ }if\text{ }-1<X_{1}\leq 0 \\ 
\exp \left( iX_{1}^{n}\right) \text{ \ \ \ \ , \ \ \ }if\text{ }0\leq
X_{1}\leq 1 \\ 
\exp \left( i\right) \text{ \ \ \ \ , \ \ \ \ }if\text{ }1\leq X_{1}<2%
\end{array}%
\right. ,  \label{11}
\end{equation}%
for $\left( X_{2},X_{3}\right) $ ranging in $\left( -1,2\right) ^{2}$, with $%
i$\ the imaginary unit, and $\tilde{d}_{\mathcal{M}}\left( a,b\right) =d_{%
\mathcal{M}}\left( a,b\right) =\left\vert a-b\right\vert $. For $n\leq m$,
we get%
\begin{eqnarray}
d^{\left( i\right) }\left( f_{n},f_{m}\right) &=&\int_{\mathcal{B}_{0}}%
\tilde{d}_{\mathcal{M}}\left( f_{n}\left( \mathbf{X}\right) ,f_{m}\left( 
\mathbf{X}\right) \right) d\left( vol\right) =  \notag \\
&=&9\int_{0}^{1}\sqrt{2\left( 1-\cos \left( X_{1}^{n}-X_{1}^{m}\right)
\right) }dX_{1}\leq  \notag \\
&\leq &9\int_{0}^{1}\left( X_{1}^{n}-X_{1}^{m}\right) dX_{1}=9\left( \frac{1%
}{n+1}-\frac{1}{m+1}\right) ,  \label{12}
\end{eqnarray}%
so that the sequence $f_{n}$ is Cauchy's but, as $n\rightarrow +\infty $,
its limit $f\left( \mathbf{X}\right) $, with%
\begin{equation}
f\left( \mathbf{X}\right) =\left\{ 
\begin{array}{c}
0\text{ \ \ , \ \ }if\text{ }-1<X_{1}<1 \\ 
\exp \left( i\right) \text{ \ \ \ \ , \ \ \ \ }if\text{ }1\leq X_{1}<2%
\end{array}%
\right. ,  \label{13}
\end{equation}%
is not continuous. The physical situation covered by such an example could
be (rather ideally) the one of a porous medium with connected voids full of
normal superfluid Helium (i.e., $^{4}$He; see [4] pag. 7) which, to the
limit (\ref{13}), percolates on one side with respect to the plane $X_{1}=1$.
\end{description}

\textbf{Remark 5}. The metric (\ref{5}) may furnish unreasonable results for
unbounded bodies. In fact, let us consider a Cartesian frame $%
0X_{1}X_{2}X_{3}$, and a domain $B=D_{q}\times \left[ 0,+\infty \right) $
where $D_{q}$ is the square $\left( -1,1\right) ^{2}$ in the plane $X_{3}=0$%
. Imagine that the closure of the domain $B$ is occupied by a semi-infinite
beam made of a two-phase material so that the order parameter is a scalar
ranging in $\left[ 0,1\right] $ and representing the volume fraction of one
phase. We consider a distribution of the two phases homogeneous with respect
to $X_{3}$ so that in a given configuration the distribution of the phases
is expressed by the order parameter mapping $B\ni \left(
X_{1},X_{2},X_{3}\right) \rightarrow \nu _{1}=\tilde{\nu}_{1}\left(
X_{1},X_{2}\right) \in \mathcal{M}\equiv \left[ 0,1\right] $. We consider
another distribution of the two phases, namely $\tilde{\nu}_{2}=1-\tilde{\nu}%
_{1}$ and select $\tilde{d}_{\mathcal{M}}\left( x,y\right) =d_{\mathcal{M}%
}\left( x,y\right) =\left\vert x-y\right\vert $ so that the integral $%
\int_{D_{q}}d_{\mathcal{M}}\left( \nu _{1},\nu _{2}\right) d\left(
area\right) =\int_{D_{q}}\left\vert 2\nu _{1}-1\right\vert d\left(
area\right) $ is a certain constant positive $K$ (when $\nu _{1}$\ is not a
constant equal to $\frac{1}{2}$), then the distance $d^{\left( i\right)
}\left( \tilde{\nu}_{1},\tilde{\nu}_{2}\right) $ reduces to the integral of $%
K$ over $\left[ 0,+\infty \right) $ which is $+\infty $. In other words, the
two configurations obtained through a rigid rotation of the beam are at
infinite distance if one uses the integral distance (\ref{5}). The same
result does not occur with the distances (\ref{6}) and (\ref{7}) and the
calculation reduces basically to the evaluation of the maximum of the
function $\left\vert 2\tilde{\nu}_{1}-1\right\vert $ varying $\left(
X_{1},X_{2}\right) $ in $D_{q}$.

\ \ \ \ \ \ \ \ \ \ 

To control the behavior of the derivatives of $\mathbf{\nu }$, basically of $%
\nabla \mathbf{\tilde{\nu}}$ with $\nabla \mathbf{\nu }\equiv \nabla \mathbf{%
\tilde{\nu}}\left( \mathbf{X}\right) \in Hom\left( T_{\mathbf{X}}\mathcal{B}%
_{0},T_{\mathbf{\nu }}\mathcal{M}\right) $, we find problematic to act
directly on $Hom\left( T\mathcal{B}_{0},T\mathcal{M}\right) $. In fact, when
we select $\mathbf{\tilde{\nu}}_{1}$ and $\mathbf{\tilde{\nu}}_{2}$, at a
given $\mathbf{X}$, we get in general $\mathbf{\nu }_{1}\neq \mathbf{\nu }%
_{2}$ so that we get $\nabla \mathbf{\nu }_{1}\in Hom\left( T_{\mathbf{X}}%
\mathcal{B}_{0},T_{\mathbf{\nu }_{1}}\mathcal{M}\right) $ and $\nabla 
\mathbf{\nu }_{2}\in Hom\left( T_{\mathbf{X}}\mathcal{B}_{0},T_{\mathbf{\nu }%
_{2}}\mathcal{M}\right) $. Though $\mathbf{g}_{\mathcal{M}}$\ induces
natural metric structures over $T^{\ast }\mathcal{M}$ and tensor product
bundles of $T\mathcal{M}$ and $T^{\ast }\mathcal{M}$, it does not give a
natural way of comparing $\nabla \mathbf{\nu }_{1}$\ and $\nabla \mathbf{\nu 
}_{2}$\ because $\nabla \mathbf{\nu }$\ is in certain sense like the
two-point tensor $\mathbf{F}$. To compare them it is then necessary to
transport $\nabla \mathbf{\nu }_{2}$ over $\mathbf{\nu }_{1}$ through a
connection or $\nabla \mathbf{\nu }_{1}$ over $\mathbf{\nu }_{2}$, i.e. to
transfer elements of $T_{\mathbf{\nu }_{2}}\mathcal{M}$ over $T_{\mathbf{\nu 
}_{1}}\mathcal{M}$. Since we do not assign any prevalent r\^{o}le to
Levi-Civita connection, even in the case in which $\mathcal{M}$ is complete
and we may connect $\mathbf{\nu }_{1}$ and $\mathbf{\nu }_{2}$ with a
geodetic, we face the circumstance that the transport is in general not
isometric, so that the comparison (defined in some way) of the two
derivatives at $\mathbf{\nu }_{1}$ may lead to a different result of the
same comparison at $\mathbf{\nu }_{2}$. Moreover, even in the case in which
one would choose to use both the comparison at $\mathbf{\nu }_{1}$ and the
one at $\mathbf{\nu }_{2}$, say, using half of their sum, one could face the
basic difficulty that the transport could be unbounded (due, e.g., to
topological features of $\mathcal{M}$) or the bound could not be uniform
over the curve connecting $\mathbf{\nu }_{1}$ with $\mathbf{\nu }_{2}$. In
this case, the possible choice to consider admissible only values over $%
\mathcal{M}$ that can be connected by curves assuring a uniformly bounded
transport could reduce too much the generality of $\mathcal{M}$.

By considering previous remarks, we follow a procedure leading to the
possibility to compare $\nabla \mathbf{\nu }_{1}$\ and $\nabla \mathbf{\nu }%
_{2}$\ in $Hom\left( T_{\mathbf{X}}\mathcal{B}_{0},T_{\mathbf{X}}^{\ast }%
\mathcal{B}_{0}\right) $.

To this aim, we consider first the adjoint of $\nabla \mathbf{\tilde{\nu}}$,
indicated with $\nabla \mathbf{\tilde{\nu}}^{\ast }$ and such that $\nabla 
\mathbf{\nu }^{\ast }\equiv \nabla \mathbf{\tilde{\nu}}^{\ast }\left( 
\mathbf{X}\right) \in Hom\left( T_{\mathbf{\nu }_{1}}^{\ast }\mathcal{M},T_{%
\mathbf{X}}^{\ast }\mathcal{B}_{0}\right) $, then we define%
\begin{equation}
\bar{d}_{\mathcal{M}}\left( \nabla \mathbf{\nu }_{1},\nabla \mathbf{\nu }%
_{2}\right) =\left\Vert \nabla \mathbf{\nu }_{1}^{\ast }\nabla \mathbf{\nu }%
_{1}-\nabla \mathbf{\nu }_{2}^{\ast }\nabla \mathbf{\nu }_{2}\right\Vert
\label{14}
\end{equation}%
where here $\left\Vert \cdot \right\Vert $ is the usual norm in $\mathbb{R}%
^{3}\otimes \mathbb{R}^{3}$, i.e. the usual norm of $3\times 3$ matrices
such that, for any $\mathbf{A}\in \mathbb{R}^{3}\otimes \mathbb{R}^{3}$, one
has $\left\Vert \mathbf{A}\right\Vert =\sqrt{tr\left( \mathbf{A}^{T}\mathbf{A%
}\right) }$.

\textbf{Remark 6}. We express $\nabla \mathbf{\tilde{\nu}}^{\ast }\nabla 
\mathbf{\tilde{\nu}}$ as $\left( \nabla \mathbf{\tilde{\nu}}^{\ast }\nabla 
\mathbf{\tilde{\nu}}\right) _{AB}=\nabla \mathbf{\tilde{\nu}}_{A}^{\ast
\alpha }\left( \mathbf{g}_{\mathcal{M}}\right) _{\alpha \beta }\nabla 
\mathbf{\tilde{\nu}}_{B}^{\beta }$ in components. In other words, the second
order tensor $\nabla \mathbf{\tilde{\nu}}^{\ast }\nabla \mathbf{\tilde{\nu}}$
\ is the pull-back in $\mathcal{B}_{0}$ of the metric $\mathbf{g}_{\mathcal{M%
}}$ in $\mathcal{M}$. As a consequence, the distance $\bar{d}_{\mathcal{M}%
}\left( \nabla \mathbf{\nu }_{1},\nabla \mathbf{\nu }_{2}\right) $ compares
values of the metric $\mathbf{g}_{\mathcal{M}}$ at two different points
(namely $\mathbf{\nu }_{1}$ and $\mathbf{\nu }_{2}$) of $\mathcal{M}$. In
fact, the scalar product $\nabla \mathbf{\nu }_{1}^{\ast }\nabla \mathbf{\nu 
}_{1}\cdot \left( d\mathbf{X}\otimes d\mathbf{X}\right) $ is the `length' of 
$d\mathbf{\nu }$ at $\mathbf{\nu }_{1}$.

With $\bar{d}_{\mathcal{M}}$ as above, we define the counterparts of the
distances (\ref{5}), (\ref{6}) and (\ref{7}).

\begin{description}
\item[4.] The first distance, indicated with $\bar{d}^{\left( i\right) }$,\
is defined by%
\begin{equation}
\bar{d}^{\left( i\right) }\left( \nabla \mathbf{\tilde{\nu}}_{1},\nabla 
\mathbf{\tilde{\nu}}_{2}\right) =\int_{\mathcal{B}_{0}}\bar{d}_{\mathcal{M}%
}\left( \nabla \mathbf{\nu }_{1},\nabla \mathbf{\nu }_{2}\right) d\left(
vol\right) .  \label{15}
\end{equation}

\item[5.] Let $\left\{ K_{n}\right\} $ be an exhaustion of $\mathcal{B}_{0}$%
, i.e. a compact cover of $\mathcal{B}_{0}$ such that $K_{n}\subset \overset{%
\circ }{K}_{n+1}$ for any $n$. The second distance, indicated with $\bar{d}%
^{\left( c\right) }$,\ is defined by%
\begin{equation}
\bar{d}^{\left( c\right) }\left( \nabla \mathbf{\tilde{\nu}}_{1},\nabla 
\mathbf{\tilde{\nu}}_{2}\right) =\dsum\limits_{n\in \mathbb{N}}\frac{1}{%
2^{-n}}\max_{\mathbf{X\in }K_{n}}\bar{d}_{\mathcal{M}}\left( \nabla \mathbf{%
\nu }_{1},\nabla \mathbf{\nu }_{2}\right) ,  \label{16}
\end{equation}

\item[6.] More simply, a third natural distance, indicated with $d^{\left(
s\right) }$,\ can be defined by%
\begin{equation}
\bar{d}^{\left( s\right) }\left( \nabla \mathbf{\tilde{\nu}}_{1},\nabla 
\mathbf{\tilde{\nu}}_{2}\right) =\sup_{\mathbf{X\in }\mathcal{B}_{0}}\bar{d}%
_{\mathcal{M}}\left( \nabla \mathbf{\nu }_{1},\nabla \mathbf{\nu }%
_{2}\right) ,  \label{17}
\end{equation}%
when $\sup_{\mathbf{X\in }\mathcal{B}_{0}}\bar{d}_{\mathcal{M}}$\ remains
bounded over $\mathcal{B}_{0}$.
\end{description}

\textbf{Remark 7}. We do not introduce any `normalization' of the distance
like the one in (\ref{4}). In fact, (\ref{15})-(\ref{17}) are calculated
over bundles whose fibers are linear spaces so that $\bar{d}_{\mathcal{M}%
}\left( \cdot ,\cdot \right) $\ displays possible properties of homogeneity
(natural over fiber spaces) while a normalization like (\ref{4}) would not.

Appropriate topologies on $\mathcal{C}_{\mathbf{\nu }}$ may then be induced
by the distances%
\begin{equation}
\tilde{d}^{\left( i\right) }\left( \mathbf{\tilde{\nu}}_{1},\mathbf{\tilde{%
\nu}}_{2}\right) =d^{\left( i\right) }\left( \mathbf{\tilde{\nu}}_{1},%
\mathbf{\tilde{\nu}}_{2}\right) +\bar{d}^{\left( i\right) }\left( \nabla 
\mathbf{\tilde{\nu}}_{1},\nabla \mathbf{\tilde{\nu}}_{2}\right) ,  \label{18}
\end{equation}%
\begin{equation}
\tilde{d}^{\left( c\right) }\left( \mathbf{\tilde{\nu}}_{1},\mathbf{\tilde{%
\nu}}_{2}\right) =d^{\left( c\right) }\left( \mathbf{\tilde{\nu}}_{1},%
\mathbf{\tilde{\nu}}_{2}\right) +\bar{d}^{\left( c\right) }\left( \nabla 
\mathbf{\tilde{\nu}}_{1},\nabla \mathbf{\tilde{\nu}}_{2}\right) ,  \label{19}
\end{equation}%
\begin{equation}
\tilde{d}^{\left( s\right) }\left( \mathbf{\tilde{\nu}}_{1},\mathbf{\tilde{%
\nu}}_{2}\right) =d^{\left( s\right) }\left( \mathbf{\tilde{\nu}}_{1},%
\mathbf{\tilde{\nu}}_{2}\right) +\bar{d}^{\left( s\right) }\left( \nabla 
\mathbf{\tilde{\nu}}_{1},\nabla \mathbf{\tilde{\nu}}_{2}\right) .  \label{20}
\end{equation}

\textbf{Remark 8}. A program based on the systematic analysis of the
properties of $\mathcal{C}_{\mathbf{\nu }}$ associated with each metric
above could deserve to be followed. In fact one could analyze systematically
the influence of the topological properties of $\mathcal{M}$ (such as
compactness) on the structure of $\mathcal{C}_{\mathbf{\nu }}$. Here, such a
program goes far from the main purposes of the present paper and we do not
follow it. However, the remarks above underline the nature of subtle
geometrical difficulties that one may encounter in going on and constitute a
reasonable starting point for further investigations. In any case, one
should take into account that when the manifold $\mathcal{M}$ is \emph{per se%
} a linear space or for reasons of modeling is embedded isometrically in a
linear space, the results in [2] and [18] apply directly to characterize the
topological properties of the space $\mathcal{C}$.

\subsection{Something more about kinematics}

Motions are sufficiently smooth curves over $\mathcal{C}$. For a given
interval of time $\left[ 0,\bar{t}\right] $, we then have mappings $\left[ 0,%
\bar{t}\right] \ni t\longmapsto \left( \mathbf{\tilde{x}}_{t},\mathbf{\tilde{%
\nu}}_{t}\right) \in \mathcal{C}$ and indicate with $\mathbf{x=\tilde{x}}%
\left( \mathbf{X},t\right) $ and $\mathbf{\nu =\tilde{\nu}}\left( \mathbf{X}%
,t\right) $ the current place at $t$ of a material element resting at $%
\mathbf{X}$ when $t=0$ and the current value of the order parameter.

With $\mathbf{\dot{x}}$ and $\mathbf{\dot{\nu}}$ we denote rates given by $%
\frac{d\mathbf{\tilde{x}}}{dt}\left( \mathbf{X},t\right) $ and $\frac{d%
\mathbf{\tilde{\nu}}}{dt}\left( \mathbf{X},t\right) $ respectively, with $%
\mathbf{\dot{\nu}}\in T_{\mathbf{\tilde{\nu}}\left( \mathbf{X},t\right) }%
\mathcal{M}$. They have counterparts $\mathbf{v}$ and $\mathbf{\upsilon }$
in the current place $\mathcal{B}$ given by%
\begin{equation}
\mathcal{B}\times \left[ 0,\bar{t}\right] \ni \left( \mathbf{x},t\right) 
\overset{\mathbf{\tilde{v}}}{\longmapsto }\mathbf{v=\tilde{v}}\left( \mathbf{%
x},t\right) \in T_{\mathbf{x}}\mathcal{B}  \label{21}
\end{equation}%
and%
\begin{equation}
\mathcal{B}\times \left[ 0,\bar{t}\right] \ni \left( \mathbf{x},t\right) 
\overset{\mathbf{\tilde{\upsilon}}}{\longmapsto }\mathbf{\upsilon =\tilde{%
\upsilon}}\left( \mathbf{x},t\right) \in T_{\mathbf{\nu }\left( \mathbf{x}%
\left( \mathbf{X},t\right) ,t\right) }\mathcal{M},  \label{22}
\end{equation}%
obtained through the mapping $\mathbf{X\longmapsto x=\tilde{x}}\left( 
\mathbf{X}\right) $ at each $t$. We have $\mathbf{\dot{x}=v}$ but $\mathbf{%
\upsilon =\dot{\nu}}+\left( grad\mathbf{\nu }\right) \mathbf{v}$ and we may
write also $\mathbf{\upsilon =\dot{\nu}}+\left( \nabla \mathbf{\nu }\right) 
\mathbf{F}^{-1}\mathbf{v=\dot{\nu}}-\left( \nabla \mathbf{\nu }\right) 
\mathbf{\dot{X}}$, where $\mathbf{\dot{X}}$\ is the material velocity $-%
\mathbf{F}^{-1}\mathbf{v}$ associated with the inverse mapping $\mathbf{X=%
\tilde{x}}^{-1}\left( \mathbf{x},t\right) $.

For the acceleration of the order parameter we have%
\begin{equation}
\ddot{\nu}^{\alpha }=\partial _{t}\dot{\nu}^{\alpha }+\breve{\Gamma}_{\beta
\gamma }^{\alpha }\dot{\nu}^{\beta }\dot{\nu}^{\gamma }  \label{25}
\end{equation}%
with%
\begin{equation}
\breve{\Gamma}_{\beta \gamma }^{\alpha }=\frac{1}{2}\mathbf{g}_{\mathcal{M}%
}^{\alpha \delta }\left( \partial _{\nu ^{\gamma }}\left( \mathbf{g}_{%
\mathcal{M}}\right) _{\beta \delta }+\partial _{\nu ^{\beta }}\left( \mathbf{%
g}_{\mathcal{M}}\right) _{\gamma \delta }-\partial _{\nu ^{\delta }}\left( 
\mathbf{g}_{\mathcal{M}}\right) _{\beta \gamma }\right)  \label{26}
\end{equation}%
Christoffel symbols relevant for $\mathcal{M}$. The expression (\ref{25})
enters the representation of possible inertial terms pertaining to the
substructure\footnote{%
Basic remarks about the influence of $\mathbf{\breve{\Gamma}}$ on inertia
are discussed in [7].}.

Let $\mathbf{g}$ be the metric in the ambient space (in general, for curved
frames we have $\mathbf{x\longmapsto g}\left( \mathbf{x}\right) $, i.e. $%
\mathbf{g}$ depends on the place), by indicating with $\mathbf{F}^{T}$ the
transpose of $\mathbf{F}$, the mapping%
\begin{equation}
\mathcal{B}_{0}\ni \mathbf{X}\longmapsto \left( \mathbf{F}^{T}\mathbf{F}%
\right) \equiv \mathbf{C}\left( \mathbf{X}\right) \in Sym^{+}\left( T_{%
\mathbf{X}}\mathcal{B}_{0},T_{\mathbf{X}}^{\ast }\mathcal{B}_{0}\right)
\end{equation}
is the pull-back of $\mathbf{g}$ through the deformation $\mathbf{\tilde{x}}$
and in coordinates we have $C_{AB}=\left( F^{T}\right) _{A}^{\ \ \
i}g_{ij}F_{B}^{j}$.

If $\mathcal{B}_{0}$\ is endowed with a not flat `material' metric $\mathbf{%
\gamma }$ (i.e. $\mathcal{B}_{0}\ni \mathbf{X}\overset{\mathbf{\tilde{\gamma}%
}}{\longmapsto }\mathbf{\gamma =\tilde{\gamma}}\left( \mathbf{X}\right) \in
Sym^{+}\left( Vec,Vec\right) $), the difference $\left( \mathbf{C-\gamma }%
\right) \left( \mathbf{X}\right) $ is twice the non-linear deformation
tensor $\mathbf{E}\left( \mathbf{X}\right) $ which measures relative changes
of lengths by using $\mathcal{B}_{0}$ as paragon setting. In an alternative
point of view, we may consider $\mathcal{B}$ as paragon setting pushing
forward $\mathbf{\gamma }$ and comparing lengths there as explained in all
treatises on non-linear elasticity of simple bodies in chapters dealing with
measures of deformation.

In the case of complex bodies the matter may be more intricate and a general
(in certain sense abstract) treatment of measures of deformation seems to be
absent (the only exception being the basic remarks in [5]). The key point is
the specific nature of $\mathbf{\nu }$. In fact, when $\mathbf{\nu }$\
represents, e.g., a microdisplacement, an independent rotation or an
independent deformation, its gradient enters the measures of deformation
together with $\mathbf{\nu }$ itself. On the contrary, when $\mathbf{\nu }$
describes a property not related strictly with changes of lengths (say in
the case in which $\mathbf{\nu }$ represents the volume fraction of a phase
in a two-phase material or the spontaneous polarization in ferroelectrics
etc.), standard $\mathbf{C}$ and $\mathbf{E}$, or their spatial
counterparts, are sufficient to measure the macroscopic deformation.

In general, we could imagine to have a map of the type%
\begin{equation}
\left( \mathbf{F,g,\nu ,}\nabla \mathbf{\nu }\right) \longmapsto \mathbf{G}%
\left( \mathbf{F,g,\nu ,}\nabla \mathbf{\nu }\right) \in Sym^{+}\left( T_{%
\mathbf{X}}\mathcal{B}_{0},T_{\mathbf{X}}^{\ast }\mathcal{B}_{0}\right)
\label{27}
\end{equation}%
with $\mathbf{G}$ a metric in $\mathcal{B}_{0}$ involving the pull-back of $%
\mathbf{g}$, and to define a deformation tensor $\mathbf{\bar{E}}\left( 
\mathbf{X}\right) $ as $\frac{1}{2}\left( \mathbf{G-\gamma }\right) \left( 
\mathbf{X}\right) $, erasing $\mathbf{\nu }$ and/or $\nabla \mathbf{\nu }$
each time in which circumstances suggest such a cancellation.

Of course, the information furnished by (\ref{27}) is rather volatile unless
we put it in a context specifying its nature for some particular
substructure.

\textbf{Remark 9}. Cosserat [10] and micromorphic ([28]) materials are well
known classes of complex bodies in which the order parameter (or its
gradient) affects the representation of the measures of deformation (see
also essential remarks in [4]). In the former class each material element is
pictured as a `small' rigid body that may undergo rotations independent of
the surrounding material elements. Of course, neighboring relative rotations
may alter lengths, in a sense described e.g. in [10] or [4]. In the latter
class one imagines that each material element may suffer independent
deformations (it is like a ball of rubber). Measures of relative
deformations may be then introduced. Less popular is the case of
microcracked bodies and we give some details about it in order to construct
an explicit example of the possible influence of the order parameter and its
gradient on the measures of deformation. In fact, when microcracks are
smeared throughout a body, the material element is pictured as a `patch' of
matter endowed with a population of microcracks that can be considered
either as planar sharp defects not interpenetrated by interatomic bonds or
as elliptic voids with non-null volume and one dimension very small with
respect to the others. If we consider \emph{frozen} the microcracks in a
given material element placed at $\mathbf{X}$ in $\mathcal{B}_{0}$, a
standard deformation $\mathbf{\tilde{x}}$ puts it (or better its centre of
mass) in the place $\mathbf{x}=\mathbf{\tilde{x}}\left( \mathbf{X}\right) $.
Now, if we allow the microcracks to deform (say without growing irreversibly
for the sake of simplicity), the centre of mass of each material element
undergoes in principle a shift\footnote{%
Here the circumstance that there is a population of microcracks in the
material element and that each microcrack is not a spheric void is crucial.}
toward a new place $\mathbf{x}^{\prime }$. If we indicate with $\mathcal{B}%
^{\prime }$ the minimal regular region containing the collection of$\ 
\mathbf{x}^{\prime }$, each one corresponding to each\textbf{\ }$\mathbf{x}$%
, we may imagine (this is a basic point of modeling) to obtain $\mathcal{B}%
^{\prime }$ from $\mathcal{B}$ by means of a sufficiently smooth mapping $%
\mathfrak{f}$ such that $\mathbf{x}^{\prime }=\left( \mathfrak{f}\circ 
\mathbf{\tilde{x}}\right) \left( \mathbf{X}\right) $ and $\mathfrak{f}\left( 
\mathcal{B}\right) =\mathcal{B}^{\prime }$. By denoting with $grad$ the
gradient with respect to $\mathbf{x}$, as before, by chain rule we get $%
\nabla \left( \mathfrak{f}\circ \mathbf{\tilde{x}}\right) \left( \mathbf{X}%
\right) =\left( \left( grad\mathfrak{f}\right) \nabla \mathbf{\tilde{x}}%
\right) \left( \mathbf{X}\right) =\mathbf{F}^{\left( m\right) }\mathbf{F}$,
where $\mathbf{F}^{\left( m\right) }=\left( grad\mathfrak{f}\right) \left( 
\mathbf{x}\right) $ is the gradient of deformation from $\mathcal{B}$ to $%
\mathcal{B}^{\prime }$, i.e., $\mathbf{F}^{\left( m\right) }=Hom\left( T_{%
\mathbf{x}}\mathcal{B},T_{\mathbf{x}^{\prime }}\mathcal{B}^{\prime }\right) $%
. If we indicate with $\mathsf{d}\left( \mathbf{X}\right) =\left( \mathfrak{f%
}\circ \mathbf{\tilde{x}}\right) \left( \mathbf{X}\right) -\mathbf{\tilde{x}}%
\left( \mathbf{X}\right) $ the displacement from $\mathbf{x}$ to $\mathbf{x}%
^{\prime }$, defined as a field over $\mathcal{B}_{0}$, since $\nabla 
\mathsf{d}\left( \mathbf{X}\right) =\left( grad\mathsf{d}_{a}\right) \mathbf{%
F}$, with $\mathsf{d}_{a}=\mathsf{d}\circ \mathbf{\tilde{x}}^{-1}$, we get
the additive decomposition $\nabla \left( \mathfrak{f}\circ \mathbf{\tilde{x}%
}\right) \left( \mathbf{X}\right) =\mathbf{F+}\nabla \mathsf{d}\left( 
\mathbf{X}\right) $. If we write $\mathbf{F}_{tot}=\mathbf{F}+\nabla \mathsf{%
d}\left( \mathbf{X}\right) $, $\mathbf{F}_{tot}$\ is the gradient of
deformation from $\mathcal{B}_{0}$\ to $\mathcal{B}^{\prime }$. So that the
right Cauchy-Green tensor $\mathbf{C}_{tot}\left( \mathbf{X}\right) =\mathbf{%
F}_{tot}^{T}\mathbf{F}_{tot}$ involves $\nabla \mathsf{d}$. Moreover, by
comparing the additive decomposition of $\mathbf{F}_{tot}$ with the
multiplicative one, namely $\mathbf{F}^{\left( m\right) }\mathbf{F}$, we
realize that $\mathbf{F}^{\left( m\right) }=\mathbf{I+}\nabla \mathsf{d}%
\left( \mathbf{X}\right) \mathbf{F}^{-1}=\mathbf{I}+grad\mathsf{d}_{a}$.\
The direct (perhaps naive) description of the kinematics of microcracked
bodies just sketched here follows [21], [26]; however, it can be obtained by
using the procedure involving the limit of bodies described in [12], [13].

\subsection{Observers}

Three geometrical environments are necessary to describe a motion and then
the configuration at a prescribed instant of a complex body: the time
interval $\left[ 0,\bar{t}\right] $ (that we could extend in principle to
the whole real line), the ambient space $\mathcal{E}^{3}$ (of course $%
\mathcal{E}^{2}$ or $\mathcal{E}^{1}$ if we deal with two- or
one-dimensional bodies), and the manifold $\mathcal{M}$.

In classical mechanics an observer is a \emph{representation} of the measure
of time and the ambient space so that changes of observers are changes of
representation.

Here, our point of view is that the notion of observer should involve all
the descriptors of the morphology (i.e. the geometry) of the material
element. Then, for us, an observer $\mathcal{O}$ is a representation of a
triplet of geometrical settings: (i) the interval of time, (ii) the ambient
space $\mathcal{E}^{3}$ and (iii) the manifold $\mathcal{M}$ of
substructural states [25]. Such a definition is a natural extension of the
standard point of view including just (i) and (ii): it takes into account
the presence of the manifold of substructural states $\mathcal{M}$ as
natural geometric ingredient in the representation of the body.

For the sake of simplicity we shall consider observers coinciding about the
measure of time.

General changes of observers, as used below, will then involve automorphism
of $\mathcal{E}^{3}$\ and the action of arbitrary Lie groups over $\mathcal{M%
}$. In this sense the notion of change of observers involves a pair of
one-parameter families of transformations indicated below.

\begin{description}
\item[A1.] $\mathbb{R}^{+}\ni s_{2}\longmapsto \mathbf{f}_{s_{2}}^{2}\in
Aut\left( \mathcal{E}^{3}\right) $, with $\mathbf{f}_{0}^{2}$ the identity%
\footnote{$Aut\left( \mathcal{E}^{3}\right) $ is the group of automorphysms
of $\mathcal{E}^{3}$.}. We put $\mathbf{f}_{0}^{2\prime }\left( \mathbf{X}%
\right) =\mathbf{v}$, where the prime denotes differentiation with respect
to $s_{2}$.

\item[A2.] A Lie group $G$, with Lie algebra $\mathfrak{g}$, acts over $%
\mathcal{M}$. If $\xi \in \mathfrak{g}$, its action over $\mathbf{\nu \in }%
\mathcal{M}$ is indicated with $\xi _{\mathcal{M}}\left( \mathbf{\nu }%
\right) $. By indicating with $\mathbf{\nu }_{g}$ the value of $\mathbf{\nu }
$ after the action of $g\in G$, if we consider a one-parameter smooth curve $%
\mathbb{R}^{+}\ni s_{3}\longmapsto g_{s_{3}}\in G$ over $G$, such that $\xi =%
\frac{dg_{s_{3}}}{ds_{3}}\left\vert _{s_{3}=0}\right. $,\ and its
corresponding curve $s_{3}\longmapsto \mathbf{\nu }_{g_{s_{3}}}$\ over $%
\mathcal{M}$, starting from a given $\mathbf{\nu }$, we have $\xi _{\mathcal{%
M}}\left( \mathbf{\nu }\right) =\frac{d}{ds_{3}}\mathbf{\nu }%
_{g_{s_{3}}}\left\vert _{s_{3}=0}\right. $. When $G$ coincides with $%
SO\left( 3\right) $, then $\xi _{\mathcal{M}}\left( \mathbf{\nu }\right)
\equiv \mathcal{A}\left( \mathbf{\nu }\right) \mathbf{\dot{q}}$, with $%
\mathcal{A}\left( \mathbf{\nu }\right) $ defined above and $\mathbf{\dot{q}%
\wedge }\in \mathfrak{so}\left( 3\right) $.
\end{description}

We specify for future use in special cases the relations characterizing
`rigid' changes of observers, i.e. the ones characterized by the
contemporary action of $SO\left( 3\right) $ on $\mathcal{M}$ and $\mathcal{E}%
^{3}$. The latter action is induced by standard isometries of point space.

Let $\left[ 0,\bar{t}\right] \ni t\longmapsto \mathbf{Q}\left( t\right) \in
SO\left( 3\right) $, with $\mathbf{Q}\left( 0\right) =$Id$_{SO\left(
3\right) }$, be a smooth curve on $SO\left( 3\right) $. Let also $\mathcal{O}
$ be an assigned representation of $\mathcal{E}^{3}$ at $t=0$. At any $t\neq
0$ we may associate an observer $\mathcal{O}_{t}^{\prime }$ obtained from $%
\mathcal{O}$ by means of $\mathbf{Q}\left( t\right) $. The transformation
from $\mathcal{O}$ to $\mathcal{O}_{t}^{\prime }$ is isometric, then a point 
$\mathbf{x}$ seen by $\mathcal{O}$ is mapped in a point $\mathbf{x}^{\prime
}=\mathbf{w}\left( t\right) +\mathbf{Q}\left( t\right) \left( \mathbf{x-x}%
_{0}\right) $, where $t\longmapsto \mathbf{w}\left( t\right) $ is an
arbitrary point valued function smooth in time and $\mathbf{x}_{0}$ a fixed
point chosen at will in space. If we calculate the rate of $\mathbf{x}%
^{\prime }$ and pull-back it in $\mathcal{O}$ writing $\mathbf{\dot{x}}%
^{\ast }$ for $\mathbf{Q}^{T}\mathbf{\dot{x}}^{\prime }$, we get the
standard reaction%
\begin{equation}
\mathbf{\dot{x}}^{\ast }=\mathbf{\dot{x}+c}\left( t\right) +\mathbf{\dot{q}%
\wedge }\left( \mathbf{x-x}_{0}\right) ,  \label{28}
\end{equation}%
characterizing a classical change of observer, where $\mathbf{c}=\mathbf{Q}%
^{T}\mathbf{w}$ is the translational velocity and $\mathbf{\dot{q}}$ the
rotational one. If the arbitrary element $\mathbf{\dot{q}\wedge }$ of the
Lie algebra $\mathfrak{so}\left( 3\right) $ acts also on $\mathcal{M}$, it
induces a rate (that we call `rigid', with the subscript $R$, just for
reminding the circumstance that it is related to a spatial rigid body
motion) given by $\mathcal{A}\mathbf{\dot{q}}$, where $\mathcal{A}\left( 
\mathbf{\nu }\right) \in Hom\left( Vec,T_{\mathbf{\nu }}\mathcal{M}\right) $%
. Classical changes of observer in a multifield setting are then
characterized by the relations (\ref{28}) and%
\begin{equation}
\mathbf{\dot{\nu}}^{\ast }=\mathbf{\dot{\nu}}+\mathcal{A}\mathbf{\dot{q}}
\label{29}
\end{equation}%
when they agree about the measure of time. Basically, if $\mathbf{\nu }_{q}$
denotes the value of $\mathbf{\nu }$ after the right action of $SO\left(
3\right) $ over $\mathcal{M}$, we have $\mathcal{A}=\frac{d\mathbf{\nu }_{q}%
}{d\mathbf{q}}\left\vert _{\mathbf{q}=0}\right. $, where $\mathbf{q}$ is a
vector connected with $\mathbf{Q}\in SO\left( 3\right) $ by the formula $%
\mathbf{Q}=\exp \left( -\mathsf{e}\mathbf{q}\right) $.with \textsf{e}
Ricci's permutation index. For example, let $\mathcal{M}$ be coincident with 
$S^{2}$ as in the case of magnetostrictive materials). For any $\mathbf{\tau 
}\in S^{2}$, we get $\mathbf{\tau }_{\mathbf{q}}=\mathbf{Q\tau }$, then $%
\mathcal{A}=-\mathbf{\tau \wedge }$.

\subsection{Relabeling}

Another transformation playing a role in the developments presented in what
follows is the \emph{relabeling} of material elements in the reference place 
$\mathcal{B}_{0}$. With the term `relabeling' we indicate a generic
transformation of $\mathcal{B}_{0}$\ induced by a $C^{1}\left( \mathcal{B}%
_{0}\right) $ point valued mapping $\mathbf{f}^{1}$ such that (i) it is 
\emph{isocoric} (i.e. it preserves volume) and (ii) establishes also a
diffeomorphism between $\mathcal{B}_{0}$ and $\mathbf{f}^{1}\left( \mathcal{B%
}_{0}\right) $. From a physical point of view, the important part of the
action of $\mathbf{f}^{1}$ relies in a sort of `permutation' of possible
defects or, better, inhomogeneities in $\mathcal{B}_{0}$. In particular, we
will consider a one-parameter family of such isocoric diffeomorphisms
defined below:

\begin{description}
\item[A3.] $\mathbb{R}^{+}\ni s_{1}\longmapsto \mathbf{f}_{s_{1}}^{1}\in
SDiff\left( \mathcal{B}_{0}\right) $, with $\mathbf{f}_{0}^{1}$ the
identity; i.e. at each $s_{1}$ we get $\mathbf{X\longmapsto f}%
_{s_{1}}^{1}\left( \mathbf{X}\right) $, with $Div\mathbf{f}_{s_{1}}^{1\prime
}\left( \mathbf{X}\right) =0$, where the prime denotes differentiation with
respect to the parameter $s_{1}$. We put $\mathbf{f}_{0}^{1\prime }\left( 
\mathbf{X}\right) =\mathbf{w}$.
\end{description}

\section{Langrangian 3+1 forms and balance equations}

The multifield theoretical analogue of the theory of elasticity for complex
bodies (i.e., bodies with material substructure) relies on rather
articulated fiber bundles.

We start by considering a fiber bundle $\pi :\mathcal{Y\rightarrow }\mathcal{%
B}_{0}\times \left[ 0,\bar{t}\right] $ such that $\pi ^{-1}\left( \mathbf{X,}%
t\right) =\mathcal{E}^{3}\times \mathcal{M}$. A generic section $\eta \in
\Gamma \left( \mathcal{Y}\right) $ of $\mathcal{Y}$\ is then $\eta \left( 
\mathbf{X,}t\right) =\left( \mathbf{X},t,\mathbf{x,\nu }\right) $. For
sufficiently smooth sections, the first jet bundle $J^{1}\mathcal{Y}$ over $%
\mathcal{Y}$ is such that $J^{1}\mathcal{Y}\ni j^{1}\left( \eta \right)
\left( \mathbf{X,}t\right) =\left( \mathbf{X},t,\mathbf{x,\dot{x},F,\nu ,%
\dot{\nu},}\nabla \mathbf{\nu }\right) $.

Up to this point the discussion has been purely geometric. No issues related
with the constitutive nature of the body and with the interactions arising
inside it have been discussed. They enter into play here: we assume that the
general body under examination is made of (non-linear) elastic material. In
this case we may associate with it the canonical Lagrangian $3+1$ form%
\begin{equation}
L:J^{1}\mathcal{Y\rightarrow \wedge }^{3+1}\left( \mathcal{B}_{0}\times %
\left[ 0,\bar{t}\right] \right) .  \label{30}
\end{equation}%
The definition of the elements of the space of $3+1$ forms $\mathcal{\wedge }%
^{3+1}\left( \mathcal{B}_{0}\times \left[ 0,\bar{t}\right] \right) $ would
require some care. In fact, $\mathcal{B}_{0}\times \left[ 0,\bar{t}\right] $
is a manifold with boundary coinciding only with $\left\{ 0\right\} \times 
\mathcal{B}_{0}\cup \left\{ \bar{t}\right\} \times \mathcal{B}_{0}$ (because 
$\mathcal{B}_{0}$ is open and coincides with the interior of its closure),
while the definition of odd forms is immediate on manifolds without
boundary. However, one is commonly interested in evaluating the variation of
the total Lagrangian%
\begin{equation}
\bar{L}\left( \mathcal{B}_{0}\right) =\int_{\mathcal{B}_{0}\times \left[ 0,%
\bar{t}\right] }\mathcal{L}d^{3}\mathbf{X}\wedge dt;
\end{equation}%
so that, in defining $L\left( j^{1}\left( \eta \right) \right) $, possible
problems related with boundary points do not play any r\^{o}le.

We assume that $L$ admits a density $\mathcal{L}$ such that%
\begin{equation}
L\left( j^{1}\left( \eta \right) \left( \mathbf{X,}t\right) \right) =%
\mathcal{L}\left( \mathbf{X},\mathbf{x,\dot{x},F,\nu ,\dot{\nu},}\nabla 
\mathbf{\nu }\right) d^{3}\mathbf{X\wedge }dt  \label{31}
\end{equation}%
with%
\begin{equation}
\mathcal{L}\left( \mathbf{X,x,\dot{x},F,\nu ,\dot{\nu},}\nabla \mathbf{\nu }%
\right) =\frac{1}{2}\rho _{0}\left\vert \mathbf{\dot{x}}\right\vert
^{2}+\rho _{0}\chi \left( \mathbf{\nu ,\dot{\nu}}\right) -\rho _{0}e\left( 
\mathbf{X,F,\nu ,}\nabla \mathbf{\nu }\right) -\rho _{0}w\left( \mathbf{%
x,\nu }\right) ,  \label{32}
\end{equation}%
where $\rho _{0}$\ is the referential mass density (conserved during the
motion), $\chi $\ the substructural \emph{kinetic co-energy} (needed when
physical circumstances prescribe substructural inertia), $e$ the elastic
energy density and $w$\ the density of the potential of external actions,
all per unit mass. The presence of $\mathbf{\nu }$\ in the list of entries
of $\chi $\ and $e$ is due to the non-trivial structure of $\mathcal{M}$. In
general the elements of $T\mathcal{M}$ cannot be separated invariantly
unless a parallelism can be found over $\mathcal{M}$. From classical
non-linear field theories we know that $\mathbf{x}$ disappears from the list
of entries of $e$ due to reasons of invariance, but $\mathbf{\nu }$\ does
not (see [4]).

We assume that $\mathcal{L}\left( \cdot \right) $\ be sufficiently smooth so
that we may find at least one section $\eta $ satisfying Euler-Lagrange
equations%
\begin{equation}
\overset{\cdot }{\overline{\partial _{\mathbf{\dot{x}}}\mathcal{L}}}%
=\partial _{\mathbf{x}}\mathcal{L}-Div\partial _{\mathbf{F}}\mathcal{L},
\label{33}
\end{equation}%
\begin{equation}
\overset{\cdot }{\overline{\partial _{\mathbf{\dot{\nu}}}\mathcal{L}}}%
=\partial _{\mathbf{\nu }}\mathcal{L}-Div\partial _{\nabla \mathbf{\nu }}%
\mathcal{L}.  \label{34}
\end{equation}

With respect to the families of transformations characterizing relabeling
and changes of observers, for the sake of brevity, we indicate with $\mathbf{%
f}^{1}$, $\mathbf{f}^{2}$ and $\mathbf{\nu }_{g}$\ the values $\mathbf{f}%
_{s_{1}}^{1}\left( \mathbf{X}\right) $, $\mathbf{f}_{s_{2}}^{2}\left( 
\mathbf{x}\right) $, $\mathbf{\nu }_{g_{s_{3}}}\left( \mathbf{X}\right) $.

\begin{definition}
A Lagrangian density $\mathcal{L}$ is invariant with respect to the action
of $\mathbf{f}_{s_{1}}^{1}$, $\mathbf{f}_{s_{2}}^{2}$ and $G$ if%
\begin{equation*}
\mathcal{L}\left( \mathbf{X,x,\dot{x},F,\nu ,\dot{\nu},}\nabla \mathbf{\nu }%
\right) =
\end{equation*}%
\begin{equation}
=\mathcal{L}\left( \mathbf{f}^{1}\mathbf{,f}^{2}\mathbf{,}\left( grad\mathbf{%
\mathbf{f}}^{2}\right) \mathbf{\dot{x},}\left( grad\mathbf{\mathbf{f}}%
^{2}\right) \mathbf{F}\left( \nabla \mathbf{f}^{1}\right) ^{-1}\mathbf{,\nu }%
_{g},\mathbf{\dot{\nu}}_{g},\left( \nabla \mathbf{\nu }_{g}\right) \left(
\nabla \mathbf{f}^{1}\right) ^{-1}\right) ,  \label{36}
\end{equation}%
for any $g\in G$ and $s_{1},s_{2}\in \mathbb{R}^{+}$.
\end{definition}

\ \ \ \ \ \ \ \ \ \ \ \ \ \ \ \ \ \ \ \ \ \ \ \ \ \ \ \ \ \ \ \ \ \ \ \ \ \
\ 

Let $\mathcal{Q}$\ and $\mathfrak{F}$\ be \emph{scalar} and \emph{vector}
densities given respectively by%
\begin{equation}
\mathcal{Q}=\partial _{\mathbf{\dot{x}}}\mathcal{L\cdot }\left( \mathbf{v}-%
\mathbf{Fw}\right) +\partial _{\mathbf{\dot{\nu}}}\mathcal{L\cdot }\left(
\xi _{\mathcal{M}}\left( \mathbf{\nu }\right) \mathbf{-}\left( \nabla 
\mathbf{\nu }\right) \mathbf{w}\right) ,  \label{37}
\end{equation}%
\begin{equation}
\mathfrak{F}=\mathcal{L}\mathbf{w+}\left( \partial _{\mathbf{F}}\mathcal{L}%
\right) ^{T}\left( \mathbf{v}-\mathbf{Fw}\right) +\left( \partial _{\nabla 
\mathbf{\nu }}\mathcal{L}\right) ^{T}\left( \xi _{\mathcal{M}}\left( \mathbf{%
\nu }\right) \mathbf{-}\left( \nabla \mathbf{\nu }\right) \mathbf{w}\right) .
\label{38}
\end{equation}

\ \ \ \ \ \ \ \ \ \ \ \ \ \ \ \ \ \ \ \ \ \ \ \ \ \ \ \ \ \ \ \ \ \ \ \ \ \ 

\textbf{Theorem 1 }[9]. \emph{If the Lagrangian density} $\mathcal{L}$ \emph{%
is invariant under} $\mathbf{f}_{s_{1}}^{1}$, $\mathbf{f}_{s_{2}}^{2}$\emph{%
\ and }$\emph{G}$\emph{, for any} $g\in G$ \emph{and} $s_{1},s_{2}\in 
\mathbb{R}^{+}$,\emph{\ then}%
\begin{equation}
\mathcal{\dot{Q}}+Div\mathfrak{F}=0\text{.}  \label{39}
\end{equation}

\ \ \ \ \ \ \ \ 

The detailed proof is contained in [9]. Here we remind only that it is based
on a direct explicit calculation of the terms in (\ref{39}) on the basis of (%
\ref{37}) and (\ref{38}), and on the exploitation of the relations%
\begin{eqnarray}
\frac{d}{ds_{1}}\mathcal{L}\left\vert _{s_{1}=0,s_{2}=0,s_{3}=0}\right. &=&0,%
\text{ \ \ }\frac{d}{ds_{2}}\mathcal{L}\left\vert
_{s_{1}=0,s_{2}=0,s_{3}=0}\right. =0,  \notag \\
\frac{d}{ds_{3}}\mathcal{L}\left\vert _{s_{1}=0,s_{2}=0,s_{3}=0}\right. &=&0,
\end{eqnarray}%
which are consequences of the request of invariance for $\mathcal{L}$.

\ \ \ \ \ \ \ \ \ \ \ \ \ \ \ \ \ \ \ 

\textbf{Corollaries} [9]. The following statements hold true for any
Lagrangian density of the type (\ref{32}) in the sense of Definition 1:

\begin{enumerate}
\item If $\mathbf{\mathbf{f}}_{s_{2}}^{2}$ alone acts on $\mathcal{L}$
leaving $\mathbf{v}$ \emph{arbitrary}, from (\ref{39}) we get Cauchy's
balance of momentum%
\begin{equation}
\rho _{0}\mathbf{\ddot{x}=}\rho _{0}\mathbf{b+}Div\mathbf{P},  \label{40}
\end{equation}%
where $\mathbf{P=-}\partial _{\mathbf{F}}\mathcal{L}\in Hom\left( T_{\mathbf{%
X}}^{\ast }\mathcal{B}_{0},T_{\mathbf{x}}^{\ast }\mathcal{B}\right) $ is the
first Piola-Kirchhoff stress and $\mathbf{b}=\partial _{\mathbf{x}}\mathcal{L%
}\in T_{\mathbf{x}}^{\ast }\mathcal{B}$ the vector of body forces.

\item If $G$ \emph{arbitrary} acts alone on $\mathcal{L}$, from (\ref{39})
and the arbitrariness of the element $\xi $\ of the Lie algebra of $G$
(chosen to define $\xi _{\mathcal{M}}\left( \mathbf{\nu }\right) $) we get
Capriz's balance of substructural interactions%
\begin{equation}
\rho _{0}\left( \overset{\cdot }{\overline{\partial _{\mathbf{\dot{\nu}}%
}\chi }}-\partial _{\mathbf{\nu }}\chi \right) =-\mathbf{z}+\rho _{0}\mathbf{%
\beta }^{ni}+Div\mathcal{S},  \label{41}
\end{equation}%
in covariant way, where $\mathbf{\beta }^{ni}=-\rho _{0}\partial _{\mathbf{%
\nu }}w\in T_{\mathbf{\nu }}^{\ast }\mathcal{M}$ represents bulk
non-inertial external interactions acting on the substructure, $\mathcal{S}%
=-\partial _{\nabla \mathbf{\nu }}\mathcal{L}\in Hom\left( T_{\mathbf{X}%
}^{\ast }\mathcal{B}_{0},T_{\mathbf{\nu }}^{\ast }\mathcal{M}\right) $ takes
into account contact substructural interactions between neighboring material
elements (and is called \emph{microstress}) and $\mathbf{z=-}\rho
_{0}\partial _{\mathbf{\nu }}e\in T_{\mathbf{\nu }}^{\ast }\mathcal{M}$
indicates self-interactions of the substructure in each material element
(and is called \emph{self-force}).

\item Let $G=SO\left( 3\right) $ and, for any element $\mathbf{\dot{q}}%
\wedge $\ of its Lie algebra, $\mathbf{\mathbf{f}}_{s_{3}}^{2}$ be such that 
$\mathbf{v=\dot{q}}\wedge \left( \mathbf{x}-\mathbf{x}_{0}\right) $ with $%
\mathbf{x}_{0}$\ a fixed point in space. In other words, we require that the
same copy of $SO\left( 3\right) $ acts both on the ambient space and on $%
\mathcal{M}$. If we require the invariance of $e$\ with respect to $SO\left(
3\right) $, we obtain%
\begin{equation}
skw\left( \partial _{\mathbf{F}}e\mathbf{F}^{T}\right) =\mathsf{e}\left( 
\mathcal{A}^{T}\partial _{\mathbf{\nu }}e+\left( \nabla \mathcal{A}%
^{T}\right) ^{t}\partial _{\nabla \mathbf{\nu }}e\right) \mathbf{,}
\label{42}
\end{equation}%
where $\mathsf{e}$ is Ricci's alternator and $skw\left( \cdot \right) $
extracts the skew-symmetric part of its argument. Previous statement render
more perspicuous Remark 3 of [9].

\item If $\mathbf{f}_{s_{1}}^{1}$ alone acts on $\mathcal{L}$, with $\mathbf{%
w}$ \emph{arbitrary}, from (\ref{39}) we get 
\begin{equation}
\overset{\cdot }{\overline{\left( \mathbf{F}^{T}\partial _{\mathbf{\dot{x}}}%
\mathcal{L}+\nabla \mathbf{\nu }^{T}\partial _{\mathbf{\dot{\nu}}}\mathcal{L}%
\right) }}-Div\left( \mathbb{P-}\left( \frac{1}{2}\rho _{0}\left\vert 
\mathbf{\dot{x}}\right\vert ^{2}+\rho _{0}\chi \left( \mathbf{\nu ,\dot{\nu}}%
\right) \right) \mathbf{I}\right) -\partial _{\mathbf{X}}\mathcal{L}=\mathbf{%
0}  \label{43}
\end{equation}%
where $\mathbb{P}=\rho _{0}e\mathbf{I}-\mathbf{F}^{T}\mathbf{P}-\nabla 
\mathbf{\nu }^{T}\mathcal{S}\in Aut\left( T_{\mathbf{X}}^{\ast }\mathcal{B}%
_{0}\right) $, with $\mathbf{I}$ the second order unit tensor, is the
modified Eshelby tensor for continua with substructure proposed in the
general setting in [22], [23]. In particular, (\ref{43}) coincides with the
balance of configurational forces for a continuum with substructure in
absence of dissipative internal forces driving defects. The balance (\ref{43}%
), in fact, is only a consequence of the invariance with respect to
relabeling.

\item Let $G=SO\left( 3\right) $ and, for any element $\mathfrak{\dot{q}}%
\wedge $\ of its Lie algebra, $\mathbf{\mathbf{f}}_{s_{1}}^{1}$ is such that 
$\mathbf{w=}\mathfrak{\dot{q}}\wedge \left( \mathbf{X}-\mathbf{X}_{0}\right) 
$ with $\mathbf{X}_{0}$ a fixed point in $\mathcal{B}_{0}$. If the material
is homogeneous, and only the special choices of $\mathbf{\mathbf{f}}%
_{s_{2}}^{2}$ and $G$ just defined act on $\mathcal{L}$, $\mathbb{P}$ is 
\emph{symmetric}.
\end{enumerate}

\textbf{Remark 9}. It may be asked whether or not an integral version of (%
\ref{41}) can be postulated as integral balance principle of substructural
interactions and then used as a first principle. In general the answer is
negative unless $\mathcal{M}$ is a linear space or is embedded in a linear
space. The reason relies upon the circumstance that the eventual integrands $%
\mathbf{\beta }$, $\mathbf{z}$ and $\mathcal{S}\mathbf{n}$\ would take
values in $T^{\ast }\mathcal{M}$ which is a non-linear space unless the
above mentioned situations of linearity for $\mathcal{M}$ occur, so the
integrals are not defined. In other words, each time we use an integral
version of (\ref{41}) we presume implicitly an embedding in a linear space;
on the contrary it does not make sense.

\section{Discontinuity surfaces}

In common special cases, solutions of (\ref{33}), (\ref{34}) or, with other
notations, of (\ref{40}), (\ref{41}) are not smooth and may display
discontinuities concentrated on submanifolds of codimension 1. Moreover,
experiments display domain formation and branching of microstructures of
various nature (see e.g. cases of nematic order in liquid crystals,
polarization in ferroelectrics, magnetization in micromagnetics,
superconducting domains in superconductors etc.).

The presence of the gradient of the morphological descriptor in the list of
entries of the energy accounts for the presence of interfaces of domain
walls in a smeared sense. However, one may ask what happens when additional
discontinuity surfaces occur and there is \emph{interaction between smeared
and sharp interfaces}. As an example, the us consider the solidification of
a two-phase flow [24]. Two types of interfaces occur in this case.
Interfaces between the two phases of the fluid and the interface between
solid and fluid parts. The two-phase fluid is a complex body, the material
element is a sort of box containing the two phases in different portions, so
that a natural morphological descriptor is the volume or mass fraction $%
\omega $\ of one phase with respect to the other. The presence of the
gradient of $\omega $ in the list of entries of the energy allows us to
account for diffuse interfaces between the two fluid phases. These diffuse
interfaces interact with the interface solid-fluid. The latter may be
described either either as a diffuse interface too or a sharp interface. The
latter interpretation is a special case of the theory discussed below.

Really, even in single phase complex materials the energy may depend on the
gradient of the order parameter to account for the inhomogeneous behavior of
substructures. For example, in the case of micromorphic materials, each
material element is a cell that may undergo micro-deformation independently
of the neighboring cells, in addition to the participation to the overall
macroscopic deformation [28]. The morphological descriptor is then a
second-order symmetric tensor representing such an additional independent
deformation. Such a situation may be representative of the behavior of
elastomers, for example. Adjacent material elements may then undergo in
principle different micro-deformations so that, in going from one element to
the other, the energetic landscape changes in accord to the gradient of the
micro-deformation and weakly non-local interaction effects of gradient type
between neighboring material elements may be accounted for. Even in this
case, additional discontinuity surfaces may occur as defects or shock and
acceleration waves. Energetic effects associated with the gradient of the
micro-deformation may influence them.

Analogous situation occurs also in quasicrystals where collective atomic
modes (phason degrees of freedom) occur \emph{within} each crystalline cell
(the material element) and are represented by a stretchable vector. The
energy depends both on the standard strain and the gradient of such a vector
[31]. Even in this case, discontinuity surfaces such as racks or
dislocations may intervene. Their behavior is influenced by the energetic
effects associated with the gradient of the order parameter (here the vector
mentioned above) but they are not strictly modeled through it.

The physical examples above allow us to clarify the point of view followed
below. \emph{We treat, in fact, situations in which smeared interfaces or
gradient effects due to inhomogeneous behavior of material substuctures
interact with sharp interfaces or, more generally, discontinuity surfaces}.

In particular, we focus our attention on a single discontinuity surface $%
\Sigma $ defined by%
\begin{equation}
\Sigma \equiv \left\{ \mathbf{X}\in cl\mathcal{B}\text{, \ }f\left( \mathbf{X%
}\right) =0\right\} ,
\end{equation}%
with $f$ a smooth function with non-singular gradient. It is oriented by the
normal vector field $\Sigma \ni \mathbf{X\longmapsto m}\left( \mathbf{X}%
\right) =\nabla f\left( \mathbf{X}\right) /\left\vert \nabla f\left( \mathbf{%
X}\right) \right\vert $ and we use the notation $\Pi $ for $\mathbf{%
I-m\otimes m}$.

For any field $\tilde{e}\left( \cdot \right) $ continuous and piecewise
continuously differentiable on $\Sigma $, we indicate with $\nabla _{\Sigma
}e$ its \emph{surface gradient} at $\mathbf{X}$, with $e=\tilde{e}\left( 
\mathbf{X}\right) $, given by $\nabla _{\Sigma }e=\nabla e\Pi $. The
opposite of the surface gradient of $\mathbf{m}$, namely $-\nabla _{\Sigma }%
\mathbf{m}$ is indicated with \textsf{L} and is the curvature tensor.

Let $\mathbf{X}\longmapsto a=\tilde{a}\left( \mathbf{X}\right) $ be a
generic field taking values in a linear space and suffering bounded
discontinuities across $\Sigma $. For $\varepsilon >0$ we indicate with $%
a^{\pm }$ the limits $\lim_{\varepsilon \rightarrow 0}a\left( \mathbf{X}\pm
\varepsilon \mathbf{m}\right) $ which are the outer ($a^{+}$) and inner ($%
a^{-}$) traces of $a$ at $\Sigma $. Then we denote with $[a]=a^{+}-a^{-}$
the jump of $a$ across $\Sigma $ and with $2\left\langle a\right\rangle
=a^{+}+a^{-}$ its average there, so that for any pair of fields $a_{1}$ and $%
a_{2}$ with the same properties of $a$ we have $[a_{1}a_{2}]=[a_{1}]\left%
\langle a_{2}\right\rangle +\left\langle a_{1}\right\rangle [a_{2}]$ if the
product $a_{1}a_{2}$\ is defined in some way and is distributive.

$[\mathbf{F}]\Pi =\mathbf{0}$ implies that $\Sigma $ is \emph{coherent}. In
other words the two pieces of the body separated by the surface do not
suffer relative shear.

Moreover, if we attribute any `virtual' motion to $\Sigma $ by means of a
vector field $\Sigma \ni \mathbf{X\longmapsto u}\left( \mathbf{X}\right) \in 
\mathbb{R}^{3}$ with normal component $U=\mathbf{u\cdot m}$ and assume that
the velocity $\mathbf{\dot{x}}$ may suffer bounded jumps across $\Sigma $,
we get the condition $\left[ \mathbf{\dot{x}}\right] =-U\left[ \mathbf{F}%
\right] \mathbf{m}$, whose proof is textbook affairs.

As essential point we assume also that the order parameter map $\mathbf{%
\tilde{\nu}}$ is \emph{continuous} across $\Sigma $. A special case in which
such an assumption plays a prominent role is Landau's theory of phase
transitions.

\subsection{The unstructured case}

We treat first the case in which $\Sigma $\ is unstructured, i.e. it is free
of own surface energy. We derive the balance equations across $\Sigma $\ in
a way that prove their covariance which has been not discussed so far. The
procedure we adopt relies on the exploitation of an integral version of the
pointwise relation (\ref{39}).

We call \emph{part} any regular subset $\mathfrak{b}$ of $\mathcal{B}_{0}$
with non vanishing volume measure. We consider an arbitrary part $\mathfrak{b%
}_{\Sigma }$ crossing $\Sigma $ in a way in which the intersection of its
boundary $\partial \mathfrak{b}_{\Sigma }$ with $\Sigma $\ be a curve and
write for it the integral counterpart of (\ref{39}), namely%
\begin{equation}
\frac{d}{dt}\int_{\mathfrak{b}_{\Sigma }}\mathcal{Q}d\left( vol\right)
+\int_{\partial \mathfrak{b}_{\Sigma }}\mathfrak{F}\cdot \mathbf{n}d\left(
area\right) =0.  \label{45}
\end{equation}%
If we take fixed the part $\mathfrak{b}_{\Sigma }$ with respect to the
virtual motion of $\Sigma $, by the use of the transport and Gauss theorems
(see, e.g., [27], [35]), as $\mathfrak{b}_{\Sigma }\rightarrow \mathfrak{b}%
_{\Sigma }\cap \Sigma $ we find%
\begin{equation}
\frac{d}{dt}\int_{\mathfrak{b}_{\Sigma }}\mathcal{Q}d\left( vol\right)
\rightarrow -\int_{\mathfrak{b}_{\Sigma }\cap \Sigma }[\mathcal{Q]}Ud\left(
vol\right) ,  \label{46}
\end{equation}%
\begin{equation}
\int_{\partial \mathfrak{b}_{\Sigma }}\mathfrak{F}\cdot \mathbf{n}d\left(
area\right) \rightarrow \int_{\mathfrak{b}_{\Sigma }\cap \Sigma }[\mathfrak{%
F]}\cdot \mathbf{m}d\left( area\right) ,  \label{47}
\end{equation}%
so that the arbitrariness of $\mathfrak{b}_{\Sigma }$\ implies the pointwise
balance%
\begin{equation}
-[\mathcal{Q]}U+[\mathfrak{F]}\cdot \mathbf{m}=0.  \label{48}
\end{equation}

\begin{proposition}
If the transformations A1, A2, A3 are smooth throughout the body, the
validity of (\ref{48}) and the invariance of $\mathcal{L}$ imply covariant
pointwise interfacial balances across $\Sigma $ as in the list below.

\begin{enumerate}
\item The action of $\mathbf{f}_{s_{2}}^{2}$ alone implies the interfacial
balance of standard interactions%
\begin{equation}
\left[ \mathbf{P}\right] \mathbf{m}=-\rho _{0}\left[ \mathbf{\dot{x}}\right]
U.  \label{49}
\end{equation}

\item The action of $G$ alone implies the interfacial balance of
substructural interactions%
\begin{equation}
\left[ \mathcal{S}\right] \mathbf{m}=-\rho _{0}\left[ \partial _{\mathbf{%
\dot{\nu}}}\chi \right] U.  \label{50}
\end{equation}

\item The action of $\mathbf{f}_{s_{1}}^{1}$ alone implies the interfacial
configurational balance along the normal $\mathbf{m}$ in absence of
dissipative forces driving $\Sigma $, namely%
\begin{equation}
\mathbf{m\cdot }\left[ \mathbb{P}\right] \mathbf{m}=\rho _{0}U\left[ \left(
\nabla \mathbf{\nu }\right) ^{\ast }\partial _{\mathbf{\dot{\nu}}}\chi %
\right] \cdot \mathbf{m}+\frac{1}{2}\rho _{0}\left[ \chi \left( \mathbf{\nu ,%
\dot{\nu}}\right) \right] -\frac{1}{2}\rho _{0}U^{2}\left[ \left\vert 
\mathbf{Fm}\right\vert ^{2}\right] .  \label{51}
\end{equation}
\end{enumerate}
\end{proposition}

The proof is a consequence of direct calculation of (\ref{48}). It is
contained in the one of Theorem 2 below.

Balance equations at discontinuity unstructured surfaces in standard
elasticity of simple bodies are derived by means of a direct evaluation of
the variation of the total Lagrangian in [15].

\subsection{The structured case}

We consider here $\Sigma $ endowed with a \emph{surface energy density} $%
\phi $ associated with surface tensions of standard and substructural nature.

Let $\mathbb{F}$ and $\mathbb{N}$ be defined by%
\begin{equation}
\mathbb{F}=\left\langle \mathbf{F}\right\rangle \Pi ,\ \ \ \ \mathbb{N}%
=\left\langle \nabla \mathbf{\nu }\right\rangle \Pi .
\end{equation}

They are \emph{surface deformation gradient} and \emph{surface gradient of
the order parameter} at $\mathbf{X}$. There exist then two mappings $\mathbb{%
\tilde{F}}$ and $\overset{\sim }{\mathbb{N}}$\ such that%
\begin{equation}
\Sigma \ni \mathbf{X}\overset{\mathbb{\tilde{F}}}{\mathbf{\longmapsto }}%
\mathbb{F=\tilde{F}}\left( \mathbf{X}\right) \in Hom\left( T_{\mathbf{X}%
}\Sigma ,T_{\mathbf{x}}\mathcal{B}\right) ,  \label{52}
\end{equation}%
\begin{equation}
\Sigma \ni \mathbf{X}\overset{\overset{\sim }{\mathbb{N}}}{\mathbf{%
\longmapsto }}\mathbb{N=}\overset{\sim }{\mathbb{N}}\left( \mathbf{X}\right)
\in Hom\left( T_{\mathbf{X}}\Sigma ,T_{\mathbf{\nu }}\mathcal{M}\right) .
\label{53}
\end{equation}%
Elementary algebra provides us%
\begin{equation}
\left\langle \mathbf{F}\right\rangle =\mathbb{F}+\left( \left\langle \mathbf{%
F}\right\rangle \mathbf{m}\right) \otimes \mathbf{m,}
\end{equation}%
\begin{equation}
\left\langle \nabla \mathbf{\nu }\right\rangle =\mathbb{N}+\left(
\left\langle \nabla \mathbf{\nu }\right\rangle \mathbf{m}\right) \otimes 
\mathbf{m.}
\end{equation}

The surface energy density is then defined by%
\begin{equation}
\left( \mathbf{m,}\mathbb{F}\mathbf{,\nu ,}\mathbb{N}\right) \overset{\tilde{%
\phi}}{\longmapsto }\phi =\tilde{\phi}\left( \mathbf{m,}\mathbb{F}\mathbf{%
,\nu ,}\mathbb{N}\right)  \label{54}
\end{equation}%
and we assume that $\tilde{\phi}$ be sufficiently smooth. It is worth noting
that the presence of the normal $\mathbf{m}$ in the list of entries of $%
\tilde{\phi}$ points out that we are considering \emph{anisotropic surfaces}.

We require the invariance of $\phi $ with respect (i) to general changes of
observers and (ii) to relabeling of $\Sigma $. As discussed above, changes
of observers are characterized by the action of the group of automorphisms
of $\mathcal{E}^{3}$ and of a generic Lie group over $\mathcal{M}$. However,
definition A3 of $\mathbf{f}_{s_{1}}^{1}$ needs to be modified in order to
describe the relabeling of $\Sigma $ in addition to the overall relabeling
of $\mathcal{B}_{0}$. Instead of $s_{1}\longmapsto \mathbf{f}_{s_{1}}^{1}$\
we should consider time-parametrized families $s_{1}\longmapsto \mathbf{\hat{%
f}}_{s_{1}}^{1}$\ of elements of $SDiff\left( \mathcal{B}_{0}\right) $
characterized by the properties listed below.

\emph{Relabeling of}\textbf{\ }$\mathcal{B}_{0}$ \emph{including} $\Sigma $.

\begin{enumerate}
\item The map $s_{1}\longmapsto \mathbf{\hat{f}}_{s_{1}}^{1}$ satisfies A1.
Moreover, we require that the field $\mathcal{B}_{0}\ni \mathbf{X}%
\longmapsto \mathbf{w=\tilde{w}}\left( \mathbf{X}\right) =\mathbf{\hat{f}}%
_{0}^{1^{\prime }}\left( \mathbf{X}\right) $ is at least of class $%
C^{1}\left( \mathcal{B}_{0}\right) $, then also across and along $\Sigma $.

\item Each $\mathbf{\hat{f}}_{s_{1}}^{1}$ preserves the elements of area of $%
\Sigma $. Namely, if $dA$ is the element of area of $\Sigma $ in $\mathcal{B}%
_{0}$, $dA=\mathbf{\hat{f}}_{s_{1}}^{1\ast }\circ dA$, where the asterisk
indicates push forward.

\item $\left( \nabla \mathbf{w}\right) \mathbf{m=0}$.

\item $\nabla _{\Sigma }w_{m}=0$, with $w_{m}=\mathbf{w\cdot m}$.
\end{enumerate}

\ \ \ \ \ \ \ \ 

Two lemmas are useful for subsequent calculations.

\ \ \ \ \ \ 

\textbf{Lemma 1}. For any isocoric vector field $\mathbf{\tilde{w}}\left( 
\mathbf{\cdot }\right) $ of class $C^{1}\left( \mathcal{B}_{0}\right) ,$%
\begin{equation}
\Pi \cdot \nabla _{\Sigma }\mathbf{w}=\left( \left( \nabla \mathbf{w}\right) 
\mathbf{m}\right) \cdot \mathbf{m}\text{,}
\end{equation}%
with $\mathbf{w=\tilde{w}}\left( \mathbf{X}\right) $.

\ \ \ \ \ \ \ \ \ \ 

The result follows from direct calculation.

\ \ \ \ \ \ \ \ \ \ \ \ 

A second order tensor field $\Sigma \ni \mathbf{X}\overset{\mathbf{\tilde{A}}%
}{\longmapsto }\mathbf{A=\tilde{A}}\left( \mathbf{X}\right) \in Hom\left( 
\mathbb{R}^{3},\mathbb{R}^{3}\right) $ is called \emph{superficial} if $%
\mathbf{Am=0}$ at each $\mathbf{X}$.

\ \ \ \ \ \ \ \ \ \ \ \ \ 

\textbf{Lemma 2}. For any second-order superficial tensor field $\mathbf{%
\tilde{A}}$ on $\Sigma $, one gets%
\begin{equation}
\mathbf{m}\cdot Div_{\Sigma }\mathbf{A}=\mathbf{A}\cdot \mathsf{L}.
\end{equation}

\ \ \ \ \ \ \ \ \ \ \ \ 

\textbf{Definition 2}. A surface energy density $\phi $ is invariant with
respect to the action of $\mathbf{\hat{f}}_{s_{1}}^{1}$, $\mathbf{f}%
_{s_{2}}^{2}$ and $G$ if%
\begin{equation}
\tilde{\phi}\left( \mathbf{m,}\mathbb{F}\mathbf{,\nu ,}\mathbb{N}\right) =%
\tilde{\phi}\left( \nabla \mathbf{\hat{f}}^{1T}\mathbf{m,}\left(
grad_{\Sigma }\mathbf{f}^{2}\right) \mathbb{F}\left( \nabla \mathbf{\hat{f}}%
^{1}\right) ^{-1}\mathbf{,\nu }_{g}\mathbf{,}\mathbb{N}_{g}\left( \nabla 
\mathbf{\hat{f}}^{1}\right) ^{-1}\right) ,
\end{equation}%
for any $g\in G$ and $s_{1},s_{2}\in \mathbb{R}^{+}$, where $\mathbb{N}%
_{g}=\left\langle \nabla \mathbf{\nu }_{g}\right\rangle \Pi $ and we have
used notations common to Definition 1.

\ \ \ \ \ \ \ \ \ \ \ \ \ \ \ \ \ \ \ \ \ \ \ \ \ 

Let $\mathfrak{X}$ be a sufficiently smooth vector density defined over $%
\Sigma $ by%
\begin{equation}
\mathfrak{X}=-\phi \Pi \mathbf{w}+\left( \partial _{\mathbb{F}}\phi \right)
^{T}\left( \mathbf{v}-\left\langle \mathbf{F}\right\rangle \mathbf{w}\right)
+\left( \partial _{\mathbb{N}}\phi \right) ^{T}\left( \xi _{\mathcal{M}%
}\left( \mathbf{\nu }\right) -\left\langle \nabla \mathbf{\nu }\right\rangle 
\mathbf{w}\right) -\left( \partial _{\mathbf{m}}\phi \otimes \mathbf{m}%
\right) \mathbf{w.}
\end{equation}%
It is the counterpart of $\mathfrak{F}$ for a surface.

\ \ \ \ \ \ \ \ \ \ \ \ \ \ \ \ \ \ \ \ \ \ \ 

\textbf{Theorem 2}. \emph{Let} $\Sigma $ \emph{be a structured surface with
surface energy} $\phi $. \emph{Let us assume}%
\begin{equation}
\frac{d}{dt}\int_{\mathfrak{b}_{\Sigma }}\mathcal{Q}d\left( vol\right)
+\int_{\partial \mathfrak{b}_{\Sigma }}\mathfrak{F}\cdot \mathbf{n}d\left(
area\right) +\int_{\partial \left( \mathfrak{b}_{\Sigma }\cap \Sigma \right)
}\mathfrak{X}\cdot \mathsf{n}d\left( length\right) =0
\end{equation}%
\emph{for any part} $\mathfrak{b}_{\Sigma }$ \emph{of} $\mathcal{B}_{0}$ 
\emph{crossing} $\Sigma $. \emph{If} $\mathcal{L}$ \emph{and} $\phi $\ \emph{%
are invariant with respect to} $\mathbf{\hat{f}}_{s_{1}}^{1}$, $\mathbf{f}%
_{s_{2}}^{2}$ \emph{and} $G$\emph{, covariant pointwise balances across} $%
\Sigma $ \emph{follow as in the list below.}

\begin{enumerate}
\item \emph{The action of }$\mathbf{f}_{s_{2}}^{2}$ \emph{alone implies the
interfacial balance of standard interactions}%
\begin{equation}
\left[ \mathbf{P}\right] \mathbf{m}+Div_{\Sigma }\mathbb{T}=-\rho _{0}\left[ 
\mathbf{\dot{x}}\right] U,  \label{GS}
\end{equation}%
\emph{where} $\mathbb{T=-}\partial _{\mathbb{F}}\phi \in Hom\left( T_{%
\mathbf{X}}\Sigma ,T_{\mathbf{x}}^{\ast }\mathcal{B}\right) $ \emph{is the
surface Piola-Kirchhoff stress}.

\item \emph{The action of} $G$ \emph{alone implies the interfacial balance
of substructural interactions}%
\begin{equation}
\left[ \mathcal{S}\right] \mathbf{m}+Div_{\Sigma }\mathbb{S}-\mathfrak{z}%
=-\rho _{0}\left[ \partial _{\mathbf{\dot{\nu}}}\chi \right] U,  \label{PMM1}
\end{equation}%
\emph{where }$\mathbb{S=-}\partial _{\mathbb{N}}\phi \in Hom\left( T_{%
\mathbf{X}}\Sigma ,T_{\mathbf{\nu }}^{\ast }\mathcal{M}\right) $ \emph{is
the surface microstress and} $\mathfrak{z}=\partial _{\mathbf{\nu }}\phi \in
T_{\mathbf{\nu }}^{\ast }\mathcal{M}$ \emph{the surface self-force}.

\item \emph{The action of} $\mathbf{\hat{f}}_{s_{1}}^{1}$ \emph{alone
implies the interfacial configurational balance along the normal} $\mathbf{m}
$ \emph{in absence of dissipative forces driving} $\Sigma $\emph{, namely}%
\begin{equation*}
\mathbf{m\cdot }\left[ \mathbb{P}\right] \mathbf{m}+\mathbb{C}_{\tan }\cdot 
\mathsf{L}+Div_{\Sigma }\mathfrak{c}=
\end{equation*}%
\begin{equation}
=\rho _{0}U\left[ \left( \nabla \mathbf{\nu }\right) ^{\ast }\partial _{%
\mathbf{\dot{\nu}}}\chi \right] \cdot \mathbf{m}+\rho _{0}\left[ \chi \left( 
\mathbf{\nu ,\dot{\nu}}\right) \right] -\frac{1}{2}\rho _{0}U^{2}\left[
\left\vert \mathbf{Fm}\right\vert ^{2}\right] ,  \label{PMM2}
\end{equation}%
\emph{where}%
\begin{equation}
\mathbb{C}_{\tan }=\phi \Pi -\mathbb{F}^{T}\mathbb{T-N}^{T}\mathbb{S}
\label{66}
\end{equation}%
\emph{is a generalized version of the surface Eshelby stress and}%
\begin{equation}
\mathfrak{c}=-\partial _{\mathbf{m}}\phi -\mathbb{T}^{T}\left\langle \mathbf{%
F}\right\rangle \mathbf{m}-\mathbb{S}^{T}\left\langle \nabla \mathbf{\nu }%
\right\rangle \mathbf{m}  \label{67}
\end{equation}%
\emph{is a surface shear}.
\end{enumerate}

By following different procedures, equation (\ref{GS}) has been derived in
[17] while equations (\ref{PMM1}) and (\ref{PMM2}) in [22], [23] (the
version of (\ref{PMM2}) for simple bodies is clearly obtained in [16]).
Here, with Theorem 2 we prove their covariance: this is the main novelty of
the theorem itself.

\subsection{Proof of Theorem 2}

\emph{Step 1}.

In accord to Definition 2, since $\tilde{\phi}$\ is invariant, we have%
\begin{equation}
\frac{d}{ds_{1}}\tilde{\phi}\left( \nabla \mathbf{\hat{f}}^{1T}\mathbf{m,}%
\left( grad_{\Sigma }\mathbf{f}^{2}\right) \mathbb{F}\left( \nabla \mathbf{%
\hat{f}}^{1}\right) ^{-1}\mathbf{,\nu }_{g}\mathbf{,}\mathbb{N}_{g}\left(
\nabla \mathbf{\hat{f}}^{1}\right) ^{-1}\right) \left\vert
_{s_{1}=0,s_{2}=0,s_{3}=0}\right. =0,
\end{equation}%
\begin{equation}
\frac{d}{ds_{2}}\tilde{\phi}\left( \nabla \mathbf{\hat{f}}^{1T}\mathbf{m,}%
\left( grad_{\Sigma }\mathbf{f}^{2}\right) \mathbb{F}\left( \nabla \mathbf{%
\hat{f}}^{1}\right) ^{-1}\mathbf{,\nu }_{g}\mathbf{,}\mathbb{N}_{g}\left(
\nabla \mathbf{\hat{f}}^{1}\right) ^{-1}\right) \left\vert
_{s_{1}=0,s_{2}=0,s_{3}=0}\right. =0,
\end{equation}%
\begin{equation}
\frac{d}{ds_{3}}\tilde{\phi}\left( \nabla \mathbf{\hat{f}}^{1T}\mathbf{m,}%
\left( grad_{\Sigma }\mathbf{f}^{2}\right) \mathbb{F}\left( \nabla \mathbf{%
\hat{f}}^{1}\right) ^{-1}\mathbf{,\nu }_{g}\mathbf{,}\mathbb{N}_{g}\left(
\nabla \mathbf{\hat{f}}^{1}\right) ^{-1}\right) \left\vert
_{s_{1}=0,s_{2}=0,s_{3}=0}\right. =0,
\end{equation}%
which correspond respectively to%
\begin{equation}
\mathbb{F}^{T}\mathbb{T\cdot \nabla }_{\Sigma }\mathbf{w}+\mathbb{N}^{T}%
\mathbb{S\cdot \nabla }_{\Sigma }\mathbf{w+\partial }_{\mathbf{m}}\phi \cdot
\left( \nabla \mathbf{w}\right) \mathbf{m}=0,  \label{NR1}
\end{equation}%
\begin{equation}
\mathbb{T\cdot \nabla }_{\Sigma }\mathbf{v}=0,  \label{NR2}
\end{equation}%
\begin{equation}
\mathfrak{z}\cdot \xi _{\mathcal{M}}\left( \mathbf{\nu }\right) +\mathbb{%
S\cdot \nabla }_{\Sigma }\xi _{\mathcal{M}}\left( \mathbf{\nu }\right) =0.
\label{NR3}
\end{equation}

They will be useful tools below.

\ \ \ \ \ \ \ \ \ \ \ \ \ \ \ \ \ 

\emph{Step 2.}

If we shrink $\mathfrak{b}_{\Sigma }$ to $\mathfrak{b}_{\Sigma }\cap \Sigma $
uniformly in time, transport and Gauss theorems (see also (\ref{46}) and (%
\ref{47})) allow us to obtain the pointwise balance%
\begin{equation}
-[\mathcal{Q]}U+[\mathfrak{F]}\cdot \mathbf{m}+Div_{\Sigma }\mathfrak{X}=0,
\label{IntNet}
\end{equation}%
thanks to the arbitrariness of $\mathfrak{b}_{\Sigma }$. Of course, the sole
difference between (\ref{IntNet}) and (\ref{48}) is the term $Div_{\Sigma }%
\mathfrak{X}$\ accounting for the interfacial structure of the surface $%
\Sigma $.

\ \ \ \ \ \ \ \ \ \ \ \ \ 

\emph{Step 3. Deduction of the referential interfacial balance of standard
interactions }(\ref{GS}).

If $\mathbf{f}^{2}$ acts alone, then%
\begin{equation}
\mathfrak{X}=\mathbb{T}^{T}\mathbf{v},\text{ \ \ }\mathcal{Q}=\rho \mathbf{%
\dot{x}\cdot v},\text{ \ \ }\mathfrak{F}=-\mathbf{P}^{T}\mathbf{v.}
\end{equation}%
Moreover, thanks to (\ref{NR2}) we get%
\begin{equation}
Div_{\Sigma }\mathfrak{X}=\mathbf{v\cdot }Div_{\Sigma }\mathbb{T}.
\end{equation}%
Then, from (\ref{IntNet}) we obtain (\ref{GS}) thanks to the arbitrariness
of $\mathbf{v}$, which is continuous across $\Sigma $.

\ \ \ \ \ \ \ \ \ \ \ \ \ \ 

\emph{Step 4. Deduction of the referential interfacial balance of
substructural interactions }(\ref{PMM1}).

If $G$ acts alone, then%
\begin{equation}
\mathfrak{X}=\mathbb{S}^{T}\xi _{\mathcal{M}}\left( \mathbf{\nu }\right) ,%
\text{ \ \ }\mathcal{Q}=\rho _{0}\partial _{\mathbf{\dot{\nu}}}\chi \mathbf{%
\cdot }\xi _{\mathcal{M}}\left( \mathbf{\nu }\right) ,\text{ \ \ }\mathfrak{F%
}=-\mathcal{S}^{T}\xi _{\mathcal{M}}\left( \mathbf{\nu }\right) \mathbf{.}
\end{equation}%
Moreover, thanks to (\ref{NR3}) we get%
\begin{equation}
Div_{\Sigma }\mathfrak{X}=\xi _{\mathcal{M}}\left( \mathbf{\nu }\right)
\cdot \left( Div_{\Sigma }\mathbb{S}-\mathfrak{z}\right) .
\end{equation}%
Then, from (\ref{IntNet}) we obtain (\ref{PMM1}) thanks to the arbitrariness
of the element $\xi $\ selected in the Lie algebra of $G$.

\ \ \ \ \ \ \ \ \ \ \ \ \ 

\emph{Step 5. Deduction of the balance of configurational forces along the
normal }$\mathbf{m}$ (\emph{i.e.} (\ref{PMM2})).

If $\mathbf{\hat{f}}^{1}$ acts alone, then%
\begin{equation}
\mathcal{Q}=-\rho \mathbf{F}^{T}\mathbf{\dot{x}\cdot w-}\rho _{0}\left(
\nabla \mathbf{\nu }\right) ^{\ast }\partial _{\mathbf{\dot{\nu}}}\chi \cdot 
\mathbf{w,}
\end{equation}%
\begin{equation}
\mathfrak{F}=\left( \left( \frac{1}{2}\rho _{0}\left\vert \mathbf{\dot{x}}%
\right\vert ^{2}+\rho _{0}\chi \left( \mathbf{\nu ,\dot{\nu}}\right) \right) 
\mathbf{I-}\mathbb{P}\right) \mathbf{w}
\end{equation}%
and, after some algebra,%
\begin{equation}
\mathfrak{X}=-\mathbb{C}_{\tan }^{T}\mathbf{w-}\mathfrak{c}w_{m},
\end{equation}%
with $\mathbb{C}_{\tan }$\ and $\mathfrak{c}$\ defined respectively by (\ref%
{66}) and (\ref{67}) and $w_{m}=\mathbf{w\cdot m}$.

Now, equation (\ref{IntNet}) comes into play: we will evaluate the component
along $\mathbf{m}$ of the right-hand side term of (\ref{IntNet}), a vector,
by taking also into account the arbitrariness of $\mathbf{w}$.

First we write%
\begin{equation*}
-[\mathcal{Q]}U+[\mathfrak{F]}\cdot \mathbf{m}=\rho _{0}[\mathbf{F}^{T}%
\mathbf{\dot{x}}]U\cdot \mathbf{w+}\rho _{0}[\left( \nabla \mathbf{\nu }%
\right) ^{\ast }\partial _{\mathbf{\dot{\nu}}}\chi ]U\cdot \mathbf{w+}
\end{equation*}%
\begin{equation}
+\frac{1}{2}\rho _{0}[\left\vert \mathbf{\dot{x}}\right\vert ^{2}]\mathbf{%
w\cdot m}+\rho _{0}[\chi \left( \mathbf{\nu ,\dot{\nu}}\right) ]\mathbf{%
w\cdot m-[}\mathbb{P}\mathbf{]w\cdot m.}  \label{C}
\end{equation}%
Moreover, by using (\ref{NR1}) and Lemma 1, we also get%
\begin{equation}
Div_{\Sigma }\left( \mathbb{C}_{\tan }^{T}\mathbf{w+}\mathfrak{c}%
w_{m}\right) =\mathbf{w\cdot }\left( Div_{\Sigma }\mathbb{C}_{\tan }+\left(
Div_{\Sigma }\mathfrak{c}\right) \mathbf{m}\right) .  \label{DivC}
\end{equation}%
Of course, in obtaining (\ref{DivC}), properties 3 and 4 of the definition
of the relabeling $\mathbf{\hat{f}}^{1}$ of $\mathcal{B}_{0}$\ including $%
\Sigma $ play a crucial role.

By inserting (\ref{C}) and (\ref{DivC}) in (\ref{IntNet}), thanks to the
arbitrariness of $\mathbf{w}$, we obtain%
\begin{equation*}
\rho _{0}[\mathbf{F}^{T}\mathbf{\dot{x}}]U+\rho _{0}[\left( \nabla \mathbf{%
\nu }\right) ^{\ast }\partial _{\mathbf{\dot{\nu}}}\chi ]U+\frac{1}{2}\rho
_{0}[\left\vert \mathbf{\dot{x}}\right\vert ^{2}]\mathbf{m+}
\end{equation*}%
\begin{equation}
+\rho _{0}[\chi \left( \mathbf{\nu ,\dot{\nu}}\right) ]\mathbf{m}=\mathbf{[}%
\mathbb{P}^{T}\mathbf{]m}+Div_{\Sigma }\mathbb{C}_{\tan }+\left( Div_{\Sigma
}\mathfrak{c}\right) \mathbf{m}  \label{SurfC}
\end{equation}%
and we shall evaluate the component along $\mathbf{m}$ of (\ref{SurfC}).

First we focus our attention on terms involving $\mathbf{\dot{x}}$\ and $%
\mathbf{\dot{\nu}}$. Let us introduce the averaged velocity $\mathbf{\bar{v}}
$ given by%
\begin{equation}
\mathbf{\bar{v}}=\left\langle \mathbf{\dot{x}}\right\rangle +U\left\langle 
\mathbf{F}\right\rangle \mathbf{m.}
\end{equation}%
With the help of the relation $[\mathbf{\dot{x}}]=-U[\mathbf{F}]\mathbf{m}$
introduced previously, we then get%
\begin{equation}
\rho _{0}[\mathbf{F}^{T}\mathbf{\dot{x}}]U\cdot \mathbf{m}=\rho _{0}[\mathbf{%
\dot{x}}]\cdot \mathbf{\bar{v}}-\rho _{0}\mathbf{[}\left\vert \mathbf{\dot{x}%
}\right\vert ^{2}\mathbf{];}  \label{Box}
\end{equation}%
in other words the normal component of the vector $\rho _{0}[\mathbf{F}^{T}%
\mathbf{\dot{x}}]U$ is equal to \emph{minus} the jump of the relative
kinetic energy $\frac{1}{2}[\rho _{0}\left\vert \mathbf{\dot{x}-\bar{v}}%
\right\vert ^{2}]$ as it is simple to verify.

Still taking into account the relation $[\mathbf{\dot{x}}]=-U[\mathbf{F}]%
\mathbf{m}$ and the definition of $\mathbf{\bar{v}}$, we also find%
\begin{equation}
\frac{1}{2}\rho _{0}[\left\vert \mathbf{\dot{x}}\right\vert ^{2}]=-\rho _{0}[%
\mathbf{\dot{x}}]\cdot \mathbf{\bar{v}+}\frac{1}{2}U^{2}[\left\vert \mathbf{%
Fm}\right\vert ^{2}].  \label{GiumChin}
\end{equation}

Now, by evaluating the normal component of (\ref{SurfC}), using (\ref{Box}),
(\ref{GiumChin}) and taking into account that $\mathbf{m\cdot }Div_{\Sigma }%
\mathbb{C}_{\tan }=\mathbb{C}_{\tan }\cdot \mathsf{L}$ as a consequence of
Lemma 2, we obtain (\ref{PMM2}) and the theorem is proven.

\section{References}

\begin{enumerate}
\item Abraham, R., Marsden, J. E., \emph{Foundations of mechanics}. The
Benjamin/Cummings Publishing Company, London, 1978.

\item Brezis, H., Li, Y. (2001), Topology and Sobolev spaces, \emph{J.
Funct. Anal.}, \textbf{183}, 321-369.

\item Capriz, G. (1985), Continua with latent microstructure, \emph{Arch.
Rational Mech. Anal.}, \textbf{90}, 43-56.

\item Capriz, G., \emph{Continua with microstructure}. Springer-Verlag,
Berlin, 1989.

\item Capriz, G. (2001), Continua with substructure. Part I and Part II, 
\emph{Phys. Mesomech.}, \textbf{3}, 5-14 and 37-50.

\item Capriz, G. (2003), Elementary preamble to a theory of granular gases, 
\emph{Rend. Sem. Mat. Univ. Padova}, \textbf{110}, 179-198.

\item Capriz, G., Giovine, P. (1997), On microstructural inertia, \emph{%
Math. Mod. meth. Appl. Sci.}, \textbf{7}, 211-216.

\item Capriz, G., Mariano, P. M. (2003), Balance at a junction in coherent
interfaces in continua with substructure, in \emph{Advances in multifield
theoris for continua with substructure}, G. Capriz and P. M. Mariano Edts,
Birkhauser, Boston, 243-263.

\item Capriz, G., Mariano, P. M. (2003), Symmetries and Hamiltonian
formalism for complex materials, \emph{J. Elasticity}, \textbf{72}, 57-70.

\item Cosserat, E., Cosserat, F., \emph{Sur la th\'{e}orie des corps
deformables}, Dunod, Paris, 1909.

\item de Gennes, P.G., Prost, J., \emph{The physics of liquid crystals},
Claredon Press (1993).

\item Del Piero, G., Owen, D. R. (1993), Structured deformation of continua, 
\emph{Arch. Rational Mech. Anal.}, \textbf{124}, 99-155.

\item Del Piero, G., Owen, D. R., \emph{Structured deformation}, Quaderni
dell'Istituto di Alta Matematica, Florence, 2000.

\item Ericksen, J. L. (1991), Liquid crystals with variable degree of
orientation, \emph{Arch. Rational Mech. Anal.}, \textbf{113}, 97-120.

\item Fatecau, R. C., Marsden, J. E., West, M. (2003), Variational
multisymplectic formulation of non-smooth continuum mechanics, in \emph{%
Perspective and problems in nonlinear sciences}, E. Kaplan, J. E. Marsden,
K. R. Sreenivasan, Springer, Berlin, 229-261.

\item Gurtin, M. E. (1995), The nature of configurational forces, \emph{%
Arch. Rational Mech. Anal.}, \textbf{131}, 67-100.

\item Gurtin, M. E., Struthers, A. (1990), Multiphase thermomechanics with
interfacial structure. III. Evolving phase boundaries in the presence of
bulk deformation, \emph{Arch. Rational Mech. Anal.}, \textbf{112}, 97-160.

\item Hang, F., Lin, F. (2003), Topology of Sobolev mappings III, \emph{%
Comm. Pure Appl. Math.}, \textbf{61}, 1383-1415.

\item Helgason, S., \emph{Differential geometry Lie groups and symmetric
spaces}. Academic Press, New York, 1978.

\item Leslie, F. M. (1968), Some constitutive equations for liquid crystals, 
\emph{Arch. Rational Mech. Anal.}, \textbf{28}, 265-283.

\item Mariano, P. M. (1999), Some remarks on the variational description of
microcracked bodies, \emph{Int. J. Non-Linear Mech.}, \textbf{34}, 633-642.

\item Mariano, P. M. (2000), Configurational forces in continua with
microstructure, \emph{Z. angew. Math. Phys.} ZAMP, \textbf{51}, 752-791.

\item Mariano, P. M. (2001), Multifield theories in mechanics of solids, 
\emph{Adv. Appl. Mech.}, \textbf{38}, 1-93.

\item Mariano, P. M. (2004), Some thermodynamical aspects of the
solidification of two-phase flows, \emph{Meccanica}, \textbf{39}, 369-382.

\item Mariano, P. M., \emph{Elements of multifield theories of complex bodies%
}, Birkhauser, Boston, in preparation.

\item Mariano, P. M., Stazi, F. L. (2004) Strain localization due to
crack-microcrack interaction: X-FEM for a multifield approach, \emph{Comp.
Meth. Appl. Mech. Eng.}, in print.

\item Marsden, J. E., Hughes, T. R. J., \emph{Mathematical foundations of
elasticity}. Prentice-Hall, New Jersey, 1983.

\item Mindlin, R. D. (1964), Micro-structures in linear elasticity, \emph{%
Arch. Rational Mech. Anal.}, \textbf{16}, 51-78.

\item Noll, W. (1958), A mathematical theory of the mechanical behavior of
continuous media. \emph{Arch. Rational Mech. Anal.}, \textbf{2}, 197-226.

\item Noll, W. (1972), A new mathematical theory of simple materials. \emph{%
Arch. Rational Mech. Anal.}, \textbf{48}, 1-50.

\item Rochal, S. B., Lorman, V. L. (2001), Minimal model of the
phonon-phason dynamics in icosahedral quasicrystls and its application to
the problem of internal friction in the \emph{i}-AlPdMn alloy, \emph{Phys.
Rev. B}, \textbf{66}, 144204 1-9.

\item Segev, R. (1994), A geometrical framework for the statics of materials
with microstructure, \emph{Mat. Models Methods Appl. Sci.}, \textbf{4},
871-897.

\item Sharpe, R. W., \emph{Differential geometry}. Springer-Verlag, Berlin,
1997.

\item \v{S}ilhav\'{y}, M. (1985), Phase transitions in non-simple bodies, 
\emph{Arch. Rational Mech. Anal.}, \textbf{88}, 135-161.

\item \v{S}ilhav\'{y}, M., \emph{The mechanics and thermodynamics of
continuous media}, Springer, Berlin, 1997.
\end{enumerate}

\end{document}